\documentclass[cernpreprint, texlive=2016, UKenglish, texmf]{atlasdoc}
 
\usepackage{atlaspackage}
\usepackage{atlasbiblatex}
 
\usepackage{atlasphysics}
\usepackage{PMGRefs-defs}
 
\graphicspath{{logos/}{figures/}}
 
\usepackage{ANA-JETM-2018-08-PAPER-defs}
 
\PreprintIdNumber{CERN-EP-2021-147}
 
\AtlasTitle{Measurement of the energy response of the ATLAS~calorimeter to charged~pions from $W^{\pm}\rightarrow\tau^{\pm}(\rightarrow\pi^{\pm}\nu_{\tau})\nu_{\tau}$ events in Run~2 data}
 
\AtlasAbstract{
The energy response of the ATLAS calorimeter is measured for single charged pions with transverse momentum in the range $10<p_\textrm{T}<300$~GeV. 
The measurement is performed using 139~$\textrm{fb}^{-1}$ of LHC proton--proton collision data at $\sqrt{s}=13$~TeV taken in Run 2 by the ATLAS detector. Charged pions originating from $\tau$-lepton decays are used to provide a sample of high-$p_\textrm{T}$ isolated particles, where the composition is known, to test an energy regime that has not previously been probed by in situ single-particle measurements. The calorimeter response to single-pions is observed to be overestimated by ${\sim}2\%$ across a large part of the $p_{\textrm{T}}$ spectrum in the central region and underestimated by ${\sim}4\%$ in the endcaps in the ATLAS simulation. The uncertainties in the measurements are ${\lesssim}1\%$ for $15<p_\textrm{T}<185$~GeV in the central region. To investigate the source of the discrepancies, the width of the distribution of the ratio of calorimeter energy to track momentum, the energies per layer and response in the hadronic calorimeter are also compared between data and simulation.
}
 
\author{The ATLAS Collaboration}
 
\AtlasRefCode{JETM-2018-08}

\AtlasJournal{Eur. Phys. J. C}
 
\AtlasJournalRef{Eur. Phys. J. C 82 (2022) 223}
 
\AtlasDOI{10.1140/epjc/s10052-022-10117-2}

\hypersetup{pdftitle={ATLAS document},pdfauthor={The ATLAS Collaboration}}
 
\newcommand*{\itpu}{\textit{pile-up}\xspace}
\newcommand*{\pu}{\text{pile-up}\xspace}
\newcommand*{\Pu}{\text{Pile-up}\xspace}
\newcommand*{\pTtrack}{\ensuremath{p_{\textrm{T}}^{\textrm{trk}}}\xspace}

\newcommand*{\EToverPT}{\ensuremath{E_{\textrm{T}}^{\textrm{EM}}/p_{\textrm{T}}^{\textrm{trk}}}\xspace}

\newcommand{\AtlasCoordFootnote}{
ATLAS uses a right-handed coordinate system with its origin at the nominal interaction point (IP)
in the centre of the detector and the $z$-axis along the beam pipe.
The $x$-axis points from the IP to the centre of the LHC ring,
and the $y$-axis points upwards.
Polar coordinates $(r,\phi)$ are used in the transverse plane,
$\phi$ being the azimuthal angle around the $z$-axis.
The pseudorapidity is defined in terms of the polar angle $\theta$ as $\eta = -\ln \tan(\theta/2)$.
Angular distance is measured in units of $\Delta R \equiv \sqrt{(\Delta\eta)^{2} + (\Delta\phi)^{2}}$.}
 
\begin{document}
 
\maketitle
 
\tableofcontents
 
\section{Introduction}
% The next lines are included from the .//sections/Intro.tex input file
The energetic proton--proton ($pp$) collisions produced by the Large Hadron Collider (LHC) predominantly result in a large number of charged and neutral hadrons which form collimated sprays known as jets. In the ATLAS Experiment~\cite{PERF-2007-01} the energy of these particles is measured using the calorimeters, and a tracking system measures the momentum of the charged particles. A particle-flow algorithm~\cite{PERF-2015-09} is used to take advantage of these two systems in the reconstruction of jets. The measurement of hadronic energy deposits is essential for the reconstruction of jets and the jet energy scale calibration relies on the accurate simulation of hadrons interacting with the calorimeter~\cite{JETM-2018-05}. A powerful method of understanding the calorimeter response to hadrons is to consider the ratio of the energy reconstructed in the calorimeter to the momentum measured in the well-aligned tracking detector~\cite{myIDTR-2019-05}. Previously, this was studied using isolated single hadrons from inclusive $pp$ collisions~\cite{PERF-2015-05, PERF-2011-05}. In this paper, the large dataset accumulated in Run 2 is exploited to select events with isolated charged pions from $\tau$-lepton decays so that a much higher energy regime can be probed.
 
The calorimeter response to electromagnetic particles such as electrons and photons is very well known due to their showers being easier to simulate accurately and precise measurements in situ using $Z\rightarrow ee$ events. The energy scale of electrons is therefore known with an uncertainty of less than 0.1\% for transverse momenta $25<\pT<70$~\GeV\ in the central region~\cite{PERF-2017-03}. The large variety of physics processes in hadronic interactions and the non-compensating nature of the ATLAS calorimeters make the response to hadrons harder to model in simulation, motivating in situ measurements using highly energetic single pions from $\tau$-lepton decays.
 
In this paper the calorimeter response across a wide range of transverse momenta is investigated. $W$~bosons are produced copiously at the LHC and can decay via $\tau$-leptons to single charged pions, providing a source of isolated pions: $W^\pm\rightarrow\tau^\pm(\rightarrow\pi^\pm\nu_\tau)\nu_\tau$. Using the Run~2 dataset recorded in 2015--2018, large samples of pions are selected in order to accurately measure the calorimeter response using the ratio of calorimeter energy to track momentum. The measured response is corrected for the contributions from the multiple $pp$ interactions per bunch crossing and the residual signals remaining in the calorimeter from adjacent bunch crossings, contributions collectively referred to as \itpu. The charged-pion response is measured as a function of track pseudorapidity, $\eta^{\textrm{trk}}$, across the tracker volume and as a function of track transverse momentum, \pTtrack, in the barrel and endcaps. The average particle energy in the highest \pTtrack bin in the endcaps is 680 (670)~\GeV\ in data (simulated data).

Previous measurements of the single-hadron response using this ratio were performed using 2010 and 2012 data~\cite{PERF-2015-05, PERF-2011-05}. These measurements focused on special runs with low instantaneous luminosity and identified isolated tracks from minimum-bias events. As high-momentum isolated tracks are not frequently produced in pure QCD interactions due to the formation of jets which consist of multiple closely spaced particles, they probed the low-momentum phase space with $p<30$~\GeV. Test-beam measurements using the Super Proton Synchrotron (SPS) probed the response to hadrons with energies, $E$, up to 350~\GeV~\cite{Gingrich:1995xe,Gingrich:1996gk,Adragna:2009zz,Dowler:2002zp,Abat:2010zza}. For the combined, electromagnetic and hadronic, barrel calorimeter, the precision of these measurements ranged from 2.8\% at $E=20$~\GeV\ to 1.4\% at $E=350$~\GeV~\cite{Abat:2010zza}. The uncertainties were dominated by the non-uniformity of the calorimeter, imperfect knowledge of the effect of the material in front of it, and, for positive beams, the fraction of protons. In this paper the response is measured in situ with a pure $\pi^\pm$ sample. 
Additionally, the latest ATLAS detector simulation with updated geometry~\cite{PERF-2015-07} and with the latest set of hadronic shower simulation models (called the \textit{physics list}~\cite{physicsList}) as used in all recent physics analyses, is used in the comparisons of data with simulation.
 
Precise knowledge of the calorimeter response at high \pT is essential for many ATLAS physics analyses. The uncertainty in the energy scale calibration of hadronic jets, the jet energy scale (JES)~\cite{JETM-2018-05}, is one of the primary experimental uncertainties in many searches for new physics and in measurements of Standard Model processes~\cite{STDM-2016-03,TOPQ-2018-05,EXOT-2018-06,SUSY-2018-22}. In situ techniques are used to derive corrections to the transverse momentum of jets to account for differences between data and simulation, as well as uncertainties in these corrections. For the highest-\pT jets in the \TeV\ regime, where in situ methods exploiting \pT balance cannot be used due to a lack of data because of the low cross-section, the determination of the uncertainties in the momentum relies on measurements of the calorimeter energy scale from single hadrons. These are based on the low-momentum in situ measurements, test-beam results and conservative extrapolations, further motivating this measurement. Also, the calibration of hadronically decaying $\tau$-leptons relies on these measurements at high \pT~\cite{PERF-2013-06}. Additionally, measurements of the internal structure of jets rely on knowledge of the calorimeter energy scale~\cite{STDM-2017-04}. Likewise, when tagging hadronically decaying heavy particles~\cite{JETM-2018-03} the jet substructure variables that are used cannot be corrected easily for calorimeter scale discrepancies, so they depend on good modelling of the underlying calorimeter energy scale.
 
The paper is organised as follows. Section~\ref{sec:detector} introduces the ATLAS detector, and Section~\ref{sec:dataandsim} describes the dataset and samples of simulated events. The selection of events for the measurement and the variable of interest are defined in Section~\ref{sec:analysis}. The energy response is evaluated in Section~\ref{sec:results} and uncertainties in the measurements are evaluated in Section~\ref{sec:systematics}. The width of the distribution of the variable of interest is measured in Section~\ref{sec:resolution} and the longitudinal segmentation of the calorimeters is exploited in Section~\ref{sec:layers} to investigate discrepancies between simulation and data. Finally, the conclusions are presented in Section~\ref{sec:conclusions}.
% End of text imported from the .//sections/Intro.tex input file
 
\section{ATLAS detector}
\label{sec:detector}
% The next lines are included from the .//sections/Detector.tex input file
The ATLAS experiment~\cite{PERF-2007-01,ATLAS-TDR-19,PIX-2018-001} at the LHC is a multipurpose particle detector with a forward--backward symmetric cylindrical geometry and a near \(4\pi\) coverage in solid angle.\footnote{\AtlasCoordFootnote} It consists of an inner tracking detector surrounded by a thin superconducting solenoid providing a \SI{2}{\tesla} axial magnetic field, electromagnetic and hadronic calorimeters, and a muon spectrometer.
 
The inner tracking detector covers the pseudorapidity range \(|\eta| < 2.5\). Radially outwards, it consists of silicon pixel, silicon microstrip, and transition radiation tracking (TRT) detectors.
 
Lead/liquid-argon (LAr) sampling calorimeters provide electromagnetic (EM) energy measurements with high granularity. A central barrel covers $|\eta|<1.475$ and endcaps cover $1.375<|\eta|<3.2$. A steel/scintillator-tile (Tile) hadron calorimeter covers the central pseudorapidity range with one barrel covering $|\eta| < 1$ and extended barrels also covering $0.8<|\eta|<1.7$. These are longitudinally segmented in shower depth into three layers. Hadronic calorimetry in the endcaps (HEC) is provided by a copper/liquid-argon calorimeter which consists of four layers. The forward regions are instrumented with LAr calorimeters for both the EM and hadronic energy measurements up to \(|\eta| = 4.9\). Additionally pre-sampler and crack scintillators recover energy lost in non-instrumented regions.
 
The muon spectrometer surrounds the calorimeters and is based on three large air-core toroidal superconducting magnets with eight coils each. The field integral of the toroids ranges between \num{2.0} and \SI{6.0}{\tesla\metre} across most of the detector. The muon spectrometer includes a system of precision tracking chambers and fast detectors for triggering.
 
A two-level trigger system is used to select events~\cite{TRIG-2019-04}. The first-level trigger is implemented in hardware and uses a subset of the detector information to accept events at a rate below \SI{100}{\kHz}. This is followed by a software-based trigger that reduces the accepted event rate to \SI{1}{\kHz} on average, depending on the data-taking conditions.
 
An extensive software suite~\cite{ATL-SOFT-PUB-2021-001} is used for real and simulated data reconstruction and analysis, for operation and in the trigger and data acquisition systems of the experiment.
% End of text imported from the .//sections/Detector.tex input file
 
\section{Data and samples of simulated events}
\label{sec:dataandsim}
% The next lines are included from the .//sections/DataAndSim.tex input file
The $pp$ collision data analysed were recorded with the ATLAS detector from 2015 to 2018 at a centre-of-mass energy, $\sqrt{s}$, of 13~\TeV. The average number of $pp$ interactions per bunch crossing was 33.7\footnote{The average number of $pp$ interactions per bunch crossing is calculated in data using the instantaneous per bunch luminosity \cite{ATLAS-CONF-2019-021} and a reference inelastic cross section which we take to be 80 mb.} and it varied from $\approx$10 to $\approx$65. Events were selected by missing transverse momentum (\met) triggers~\cite{TRIG-2019-01}. After the application of data-quality requirements~\cite{DAPR-2018-01}, the data sample corresponds to an integrated luminosity of 139~\ifb\ with an uncertainty of 1.7\%~\cite{ATLAS-CONF-2019-021} obtained using the LUCID-2 detector~\cite{LUCID2} for the primary luminosity measurements.
 
Samples of simulated events created using Monte Carlo techniques are used to model the Standard Model processes. The main signal processes are $W(\rightarrow\tau\nu_\tau)+$jets and top pair production (\ttbar), with smaller signal contributions from single top production and small background contributions from $W(\rightarrow e\nu_e,\mu\nu_\mu)+$jets and $Z(\rightarrow\nu\nu)+$jets. Other processes are cross-checked with simulated event samples and found to be negligible.
 
The production of $W+$jets and $Z+$jets was simulated with the \SHERPA[2.2.1]~\cite{Gleisberg:2008ta,Bothmann:2019yzt} generator using matrix elements (MEs) with next-to-leading-order (NLO) accuracy for up to two jets, and matrix elements with leading-order accuracy for up to four jets, calculated with the Comix~\cite{Gleisberg:2008fv} and \OPENLOOPS~\cite{Cascioli:2011va,Denner:2016kdg} libraries. They were matched with the \SHERPA parton shower~\cite{Schumann:2007mg} using the \MEPSatNLO prescription~\cite{Hoeche:2011fd,Hoeche:2012yf,Catani:2001cc,Hoeche:2009rj} with the set of tuned parameters (tune) developed by the \SHERPA authors. All polarisation effects in $\tau$-lepton production were retained by using a full matrix element calculation, and were propagated to its decay in the \textsc{HADRONS++} module, which implements intermediate hadronic resonances and spin-correlation effects. The \NNPDF[3.0nnlo] set of parton distribution functions (PDFs)~\cite{Ball:2014uwa} was used and the samples are normalised to a next-to-next-to-leading-order (NNLO) prediction~\cite{Anastasiou:2003ds}.
 
The production of \ttbar and single-top-quark events was modelled using the \POWHEGBOX[v2]~\cite{Frixione:2007nw,Nason:2004rx,Frixione:2007vw,Alioli:2010xd} generator at NLO with the \NNPDF[3.0nlo]~\cite{Ball:2014uwa} set of PDFs. These events were processed with \PYTHIA[8.230]~\cite{Sjostrand:2014zea} using the A14 tune~\cite{ATL-PHYS-PUB-2014-021} and the \NNPDF[2.3lo] set of PDFs~\cite{Ball:2012cx}. The decays of bottom and charm hadrons were modelled using the \EVTGEN[1.6.0]~\cite{EvtGen} program. In \ttbar events the \hdamp\ parameter\footnote{The \hdamp\ parameter is a resummation damping factor and one of the parameters that controls the matching of \powheg matrix elements to the parton shower and thus effectively regulates the high-\pt\ radiation against which the \ttbar\ system recoils.} was set to 1.5\,\mtop~\cite{ATL-PHYS-PUB-2016-020}. The \ttbar process is normalised to the inclusive cross-section calculation at NNLO in QCD including the resummation of next-to-next-to-leading logarithmic (NNLL) soft-gluon terms from \TOPpp[2.0]~\cite{Beneke:2011mq,Cacciari:2011hy,Baernreuther:2012ws,Czakon:2012zr,Czakon:2012pz,Czakon:2013goa,Czakon:2011xx}. The single-top-quark production processes are normalised to the inclusive cross-sections calculated at NLO in QCD with NNLL soft-gluon corrections~\cite{Aliev:2010zk,Kant:2014oha}. For single top quark production in the $Wt$-channel events, the diagram removal scheme~\cite{Frixione:2008yi} was used to remove overlap with \ttbar\ production. In all samples the top quark mass is 172.5~\GeV.
 
To study the generator dependence of the analysis an alternative sample of QCD $W$+jets events was simulated by \mgamc\!2.2.2~\cite{Alwall:2014hca} using LO-accurate MEs with up to four final-state partons and the \NNPDF[3.0nlo] set of PDFs. The events were interfaced to \PYTHIA[8.186]~\cite{Sjostrand:2007gs}, with the A14 tune using the \NNPDF[2.3lo] PDF set, for the modelling of the parton shower, hadronisation, and  underlying event. The overlap between matrix element and parton shower emissions was removed using the CKKW-L merging procedure~\cite{Lonnblad:2001iq,Lonnblad:2011xx}. The decays of bottom and charm hadrons were performed by \EVTGEN[1.2.0]~\cite{EvtGen}.
 
All samples were passed through a detailed simulation of the ATLAS detector~\cite{SOFT-2010-01} based on \GEANT~\cite{Agostinelli:2002hh}. The modelling of hadron interactions was done by the \texttt{FTFP\_BERT\_ATL}~\cite{physicsList} physics list. In this physics list the \textit{Bertini cascade model}~\cite{Guthrie:1968ue} is used for hadrons with energy below 12~\GeV, and the \textit{Fritiof string model}~\cite{ANDERSSON1987289,NILSSONALMQVIST1987387} for hadrons with energy above 9~\GeV. The probability of using each model changes smoothly across the region of overlap.
 
Furthermore, simulated inclusive inelastic $pp$ collisions were overlaid to model additional \pu collisions in the same and neighbouring bunch crossings. These were generated with \PYTHIA[8.210]~\cite{Sjostrand:2014zea} with the A3 tune~\cite{ATL-PHYS-PUB-2016-017} and \NNPDF[2.3lo] PDF set.
% End of text imported from the .//sections/DataAndSim.tex input file
 
\section{Event selection, observables and response extraction}
\label{sec:analysis}
% The next lines are included from the .//sections/Analysis.tex input file
Events are selected using the properties of reconstructed objects to obtain a high-purity sample of $W^\pm\rightarrow\tau^\pm(\rightarrow\pi^\pm\nu_\tau)\nu_\tau$ decays which are used to measure the calorimeter response. The calorimeter energy associated with the track is summed and a fit to the energy divided by the track momentum is performed to extract the calorimeter response. The reconstructed objects used to form the variable of interest are described as follows:
\begin{itemize}
\item Charged-pion tracks are reconstructed by an iterative track-finding algorithm seeded by measurements in the silicon layers of the inner detector~\cite{PERF-2015-08,ATL-PHYS-PUB-2015-051}. The precise alignment of the inner detector ensures that these tracks are accurately reconstructed with a residual sagitta bias and momentum scale bias of less than 0.1~\TeV${}^{-1}$ and $0.9\times10^{-3}$ respectively~\cite{myIDTR-2019-05}.
\item \textit{Calorimeter topoclusters} are clusters of connected calorimeter cells throughout both the EM and hadronic calorimeters. They are seeded from cells with reconstructed energy significantly above the noise~\cite{PERF-2014-07}. Their energy is the sum of the energies of the constituent cells, which are calibrated at the electromagnetic scale. This scale is defined such that the response to electromagnetic showers is correctly calibrated in all calorimeters. Due to the non-compensating nature of the ATLAS calorimeters, the response to hadronic showers is expected to be lower such that they will be under-calibrated. Single particles can form multiple topoclusters~\cite{PERF-2015-09}, so when measuring the response all clusters in both the EM and hadronic calorimeters within a region are summed.
\end{itemize}
Additionally, various other reconstructed particles and observables are used in the event selection:
\begin{itemize}
\item Electrons are reconstructed from tracks and calorimeter energy deposits. They are identified through a log-likelihood discriminant based on the track properties and the shower shape in the calorimeter. Electrons are identified with the \textit{Loose} working point and require $\pT>10$~\GeV\ and $|\eta|<2.47$~\cite{EGAM-2018-01}. 
\item Muons are reconstructed using combined fits of inner detector and muon spectrometer tracks and are required to meet the \textit{Medium} identification criteria and have $\pT>10$~\GeV\ and $|\eta|<2.7$~\cite{MUON-2018-03}. The momentum scale and resolution for muons are precisely calibrated and measured using the $Z$-boson and $J/\psi$ resonances~\cite{PERF-2015-10}.
\item Jets are reconstructed for the purpose of building the missing transverse momentum, \met{}. Particle-flow objects~\cite{PERF-2015-09} are clustered using the anti-$k_t$ algorithm~\cite{Cacciari:2008gp} with radius parameter $R=0.4$ using the FastJet package~\cite{Fastjet} and are calibrated to the scale of jets built from the momentum of the stable interacting particles entering the simulation using the same algorithm. During the calibration sequence, corrections are applied to mitigate the effects of \pu, and jets in data are corrected for residual differences between data and simulation measured in situ using various \pT-balance techniques~\cite{JETM-2018-05}. Calibrated jets are required to have $\pT>20$~\GeV. Jets originating from \pu are rejected using inner-detector information~\cite{PERF-2014-03}.
\item \met{} is reconstructed from the vector sum of the momenta of the \textit{hard objects} described above and a \textit{soft term} formed from tracks matched to the primary vertex (PV)\footnote{The primary vertex is the reconstructed vertex~\cite{PERF-2015-01} with the highest summed \pTsq of tracks associated with it.} of interest but not associated with any \textit{hard object}~\cite{PERF-2016-07}.
\end{itemize}
 
The event selection is based on the above reconstructed particles and observables, and is designed to obtain a high-purity sample of $W^\pm\rightarrow\tau^\pm(\rightarrow\pi^\pm\nu_\tau)\nu_\tau$ events with the following properties:
\begin{enumerate}
\item $\met{}>150$~\GeV\ in addition to being selected by the missing transverse momentum trigger.
\item A leading track with $\pT>10$~\GeV\ and $|\eta|<2.5$ satisfying the following criteria:
\begin{itemize}
\item The sum of the \pT of other tracks, matched to the PV, within a cone of size $\Delta R=0.3$ around the track is less than 2~\GeV.
\item The track's longitudinal impact parameter, measured relative to the PV, is less than 1.5~mm.
\item The track's transverse impact parameter, measured relative to the beamline, is less than 0.5~mm.
\item The track satisfies the \textit{tight} criteria for hits in the silicon pixel and microstrip detectors~\cite{ATLAS-CONF-2012-042}.
\item The $\chi^2$ per degree of freedom of the track fit is less than 1.5.
\item Several TRT hits are used in the track fit if the track is within $|\eta|<2.0$.
\end{itemize}
\item There are no reconstructed electrons or muons within $\Delta R=0.4$ of the leading track.
\item At least 15\% of the total energy of all topoclusters within $\Delta R=0.15$ of the track is deposited in the hadronic calorimeters.
\item There are less than two muon spectrometer track segments~\cite{MUON-2018-03} within $\Delta R=0.15$ of the track.
\item The track does not point at any region of the Tile calorimeter that had defects during that time period.
\item To reduce backgrounds where the reconstructed track is not from $W^\pm\rightarrow\tau^\pm(\rightarrow\pi^\pm\nu_\tau)\nu_\tau$, the transverse mass\footnote{The transverse mass is defined by $m_{\textrm{T}}=\sqrt{2\pTtrack\met{}[1-\cos(\Delta\phi)]}$, where $\Delta\phi$ is the angle in the azimuthal plane between the \met{} direction and the track.} of the leading track and the \met{} satisfies $m_\textrm{T}<100$~\GeV, or $m_\textrm{T}<90$~\GeV\ if $\pTtrack<30$~\GeV, or $m_\textrm{T}<80$~\GeV\ if $\pTtrack<20$~\GeV.
\item The transverse impact parameter of the track relative to the beamline is greater than: 0.04~mm if $\pTtrack<16$~\GeV, 0.03~mm if $16<\pTtrack<20$~\GeV, and 0.02~mm if $20<\pTtrack<30$~\GeV. No criteria is applied to tracks with $\pTtrack>30$~\GeV. This exploits the lifetime of the $\tau$-lepton to reduce backgrounds where the track is from a QCD process.
\end{enumerate}
The first criterion ensures that the selected events are in the well-understood region of the \met{} trigger efficiency where it is not rapidly changing with respect to \met{}. The tracking criteria select isolated tracks associated with the primary vertex that have a large number of hits to measure the momentum. The TRT hit requirements and the criterion for the $\chi^2$ per degree of freedom of the track fit ensure that the track reconstruction uses the large lever arm within the TRT volume and that the track fit is of good quality. The third, fourth and fifth criteria reject tracks that are formed from electrons, muons and converted photons. The final two criteria reject events where the track does not originate from a $\tau$-lepton; such backgrounds are only significant at low \pTtrack. They utilise the expected displacement of the track owing to the $\tau$-lepton lifetime and the expected upper bound on the transverse mass at the $W$~boson mass.
 
After this selection the sample has a high purity of $W^\pm\rightarrow\tau^\pm(\rightarrow\pi^\pm+\nu_\tau)\nu_\tau$ events with the most significant background being $W^\pm\rightarrow\tau^\pm(\rightarrow\pi^\pm+n\pi^0+\nu_\tau)\nu_\tau$ events. For each track the transverse energy of the clusters within a fixed cone with $\Delta R=0.15$ of the track, taking into account the effect of the magnetic field, is summed. The ratio of this summed transverse energy of all clusters, in all calorimeters, at the electromagnetic scale to the track \pT forms the variable of interest, \EToverPT. Based on the simulation, it is not expected that the event selection will bias the response significantly and the selection criteria are varied to check that the results of the analysis are not sensitive to the specific selection criteria.
 
The summed calorimeter energy deposit is corrected for \pu contributions. Following the methodology used in the calibration of jets, the measured median \pT density in the $y$--$\phi$ plane of the event, $\rho$~\cite{Cacciari:2007fd,JETM-2018-05}, scaled by the area of the cone within which we sum clusters, $A=\pi\times0.15^2$, is subtracted. After this subtraction there is a small residual \pu dependence which is subtracted using the average number of interactions per bunch crossing, $\mu^{\textrm{PU}}$, and the gradient, $\alpha$, determined from simulation as a function of $|\eta|$: $\alpha(|\eta|)\times\mu^{\textrm{PU}}$. Only one residual correction is applied compared to two in the calibration of jets as the \pu contributions are much smaller in this case due to the smaller area considered. This methodology is different from the previous single-hadron response measurement in a low \pu environment~\cite{PERF-2015-05, PERF-2011-05}, where \pu and physics backgrounds were both subtracted. In this analysis the \pu is subtracted but the physics background is included in the functional fit.
 
Events are selected according to their \pTtrack and $|\eta^{\textrm{trk}}|$ values to probe the different regions of phase space. Figure~\ref{fig.fitExample} shows the distribution of the variable of interest, \EToverPT, for six of the (\pTtrack, $|\eta^{\textrm{trk}}|$) bins. Due to differences in the \EToverPT scale and resolution between simulation and data, some clear discrepancies are seen in the data-to-MC ratio.
 
\begin{figure}
\centering
\includegraphics[width=0.43\textwidth]{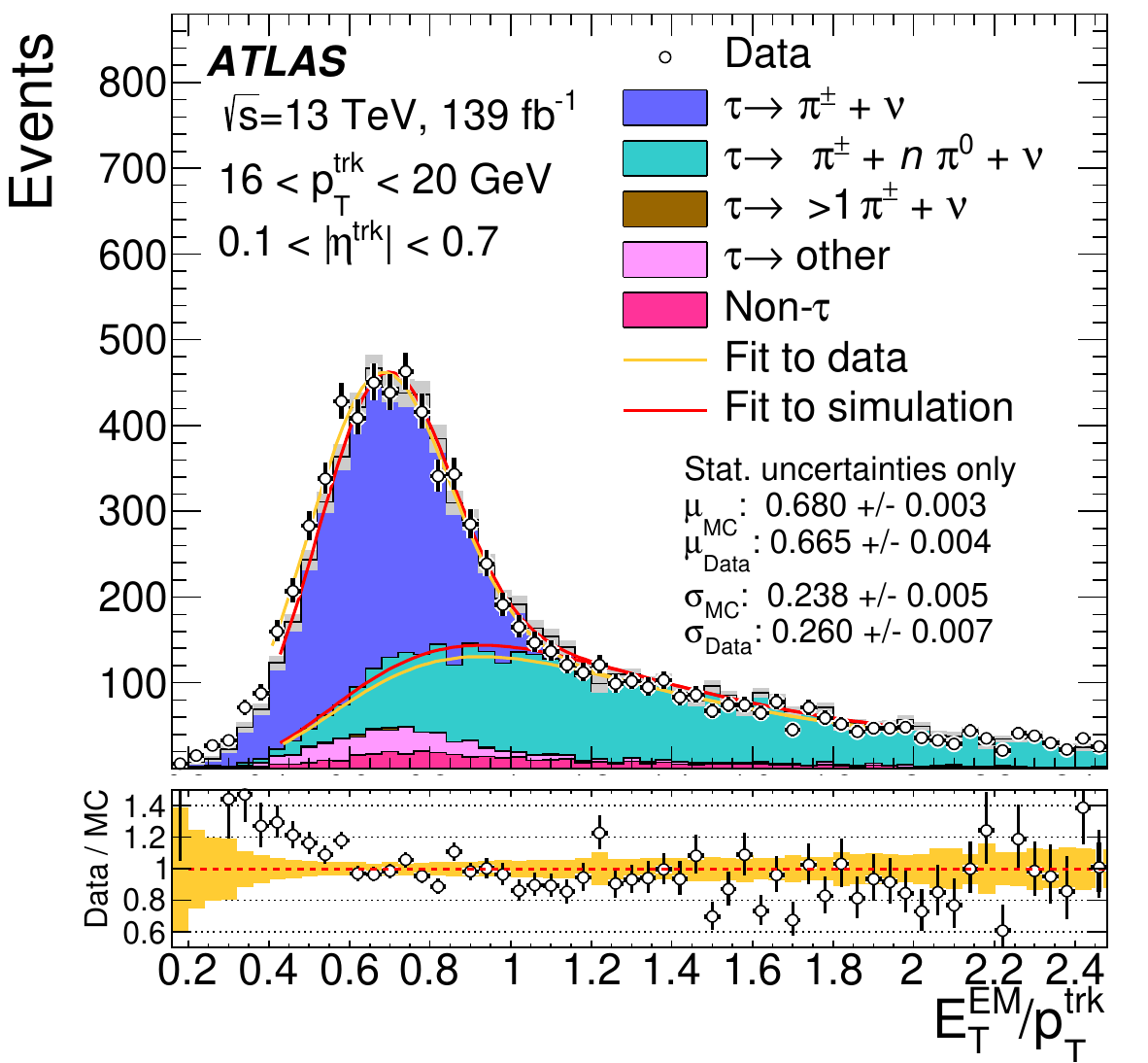}
\includegraphics[width=0.43\textwidth]{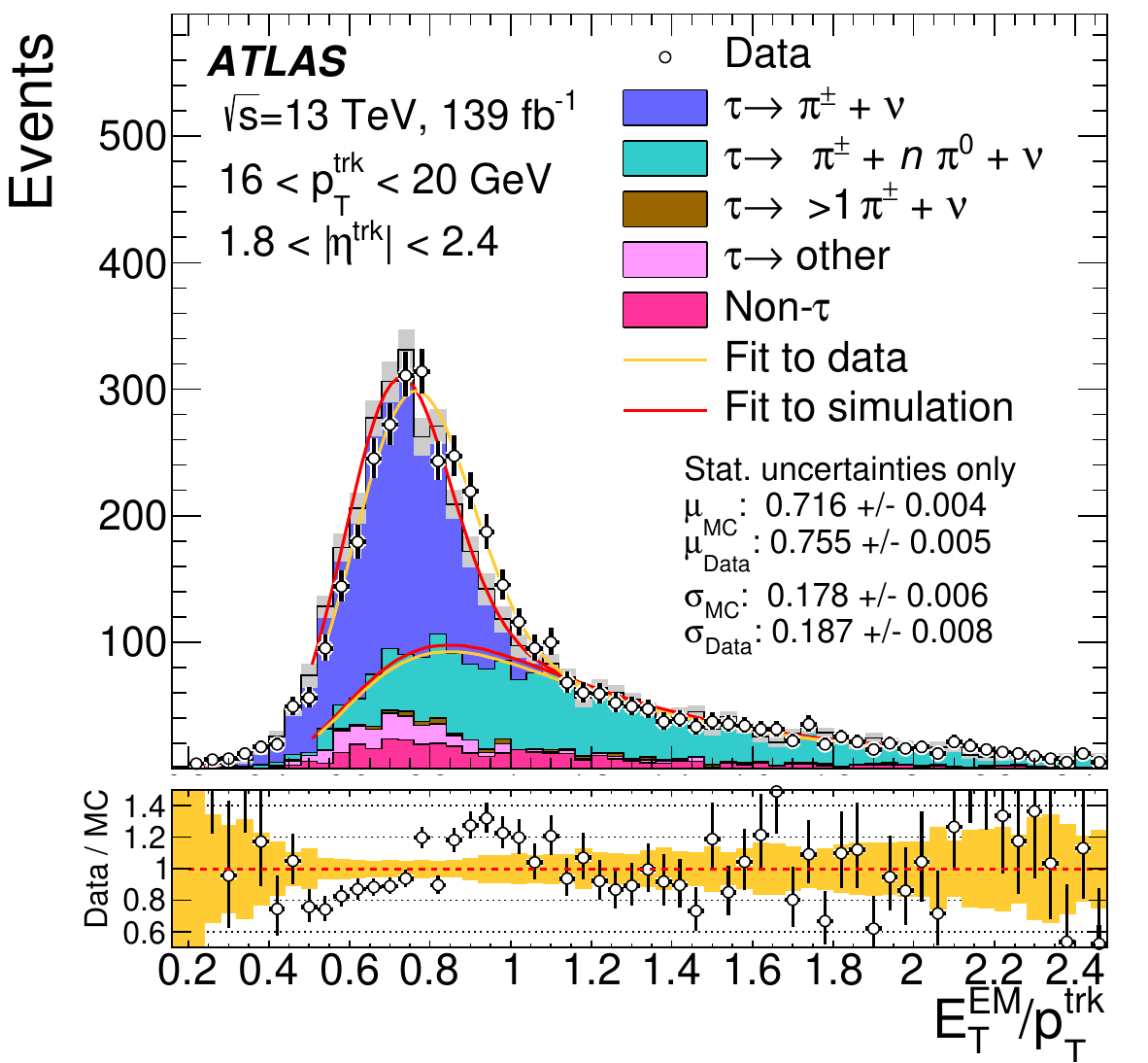}\\
\includegraphics[width=0.43\textwidth]{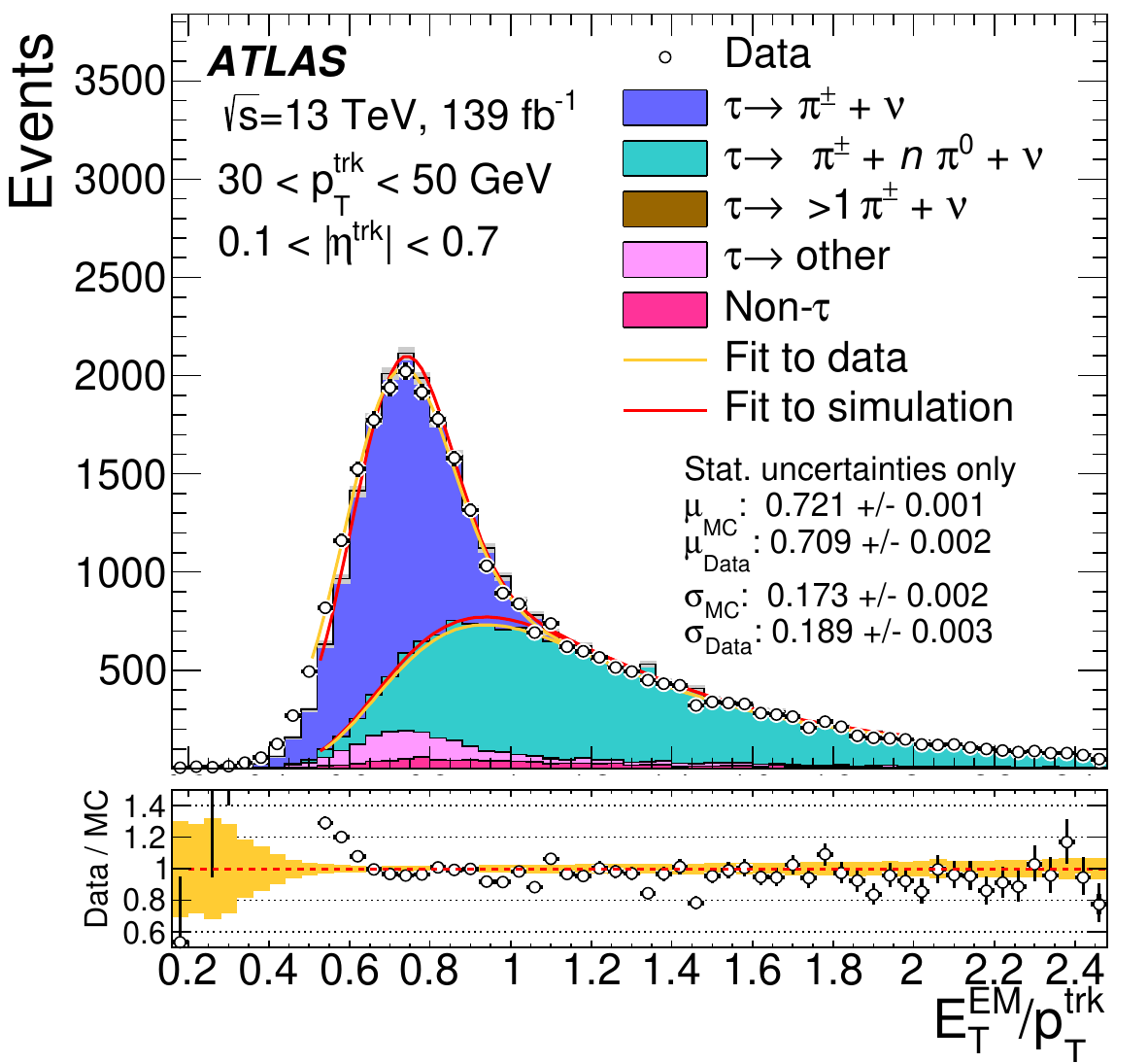}
\includegraphics[width=0.43\textwidth]{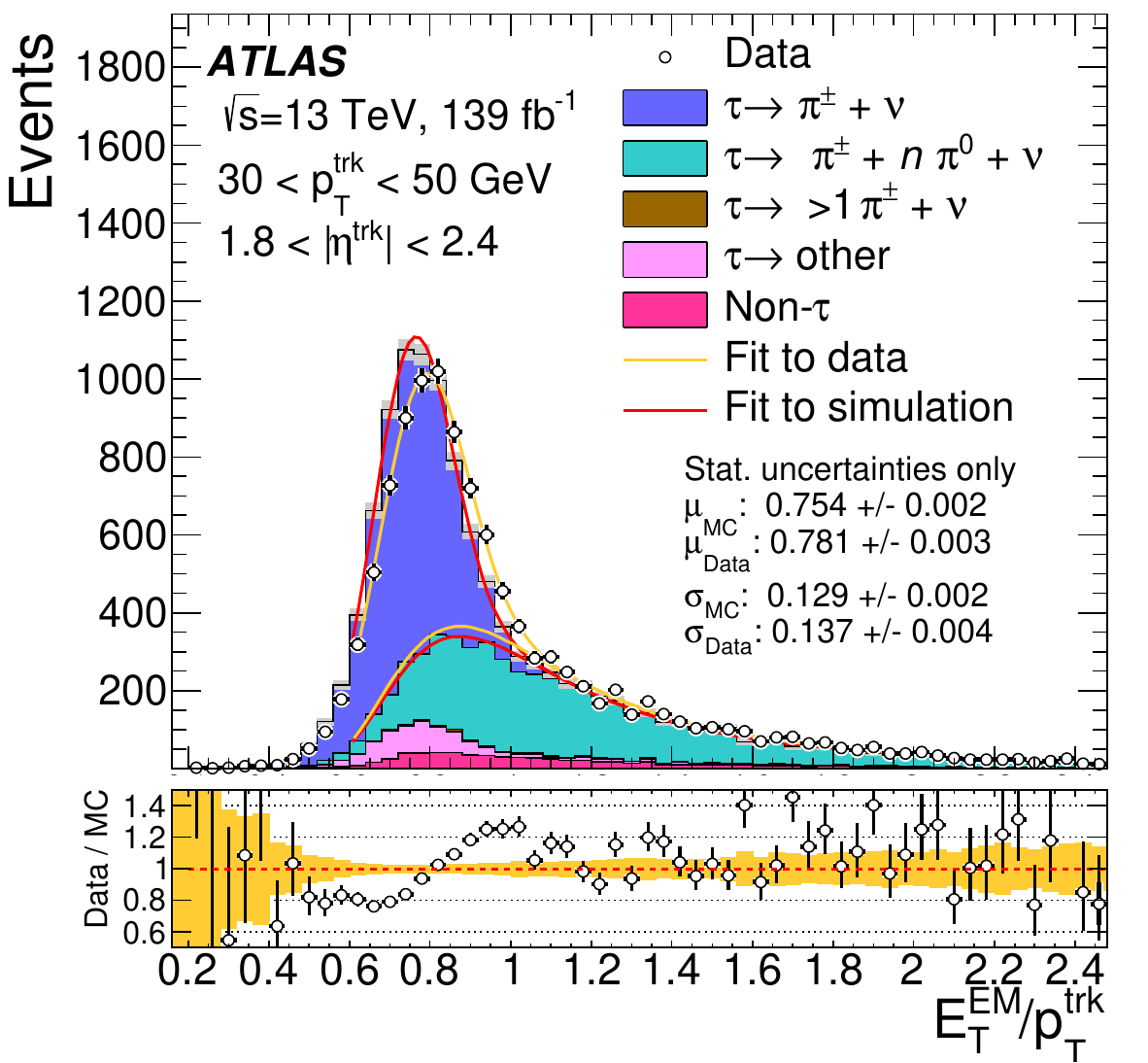}\\
\includegraphics[width=0.43\textwidth]{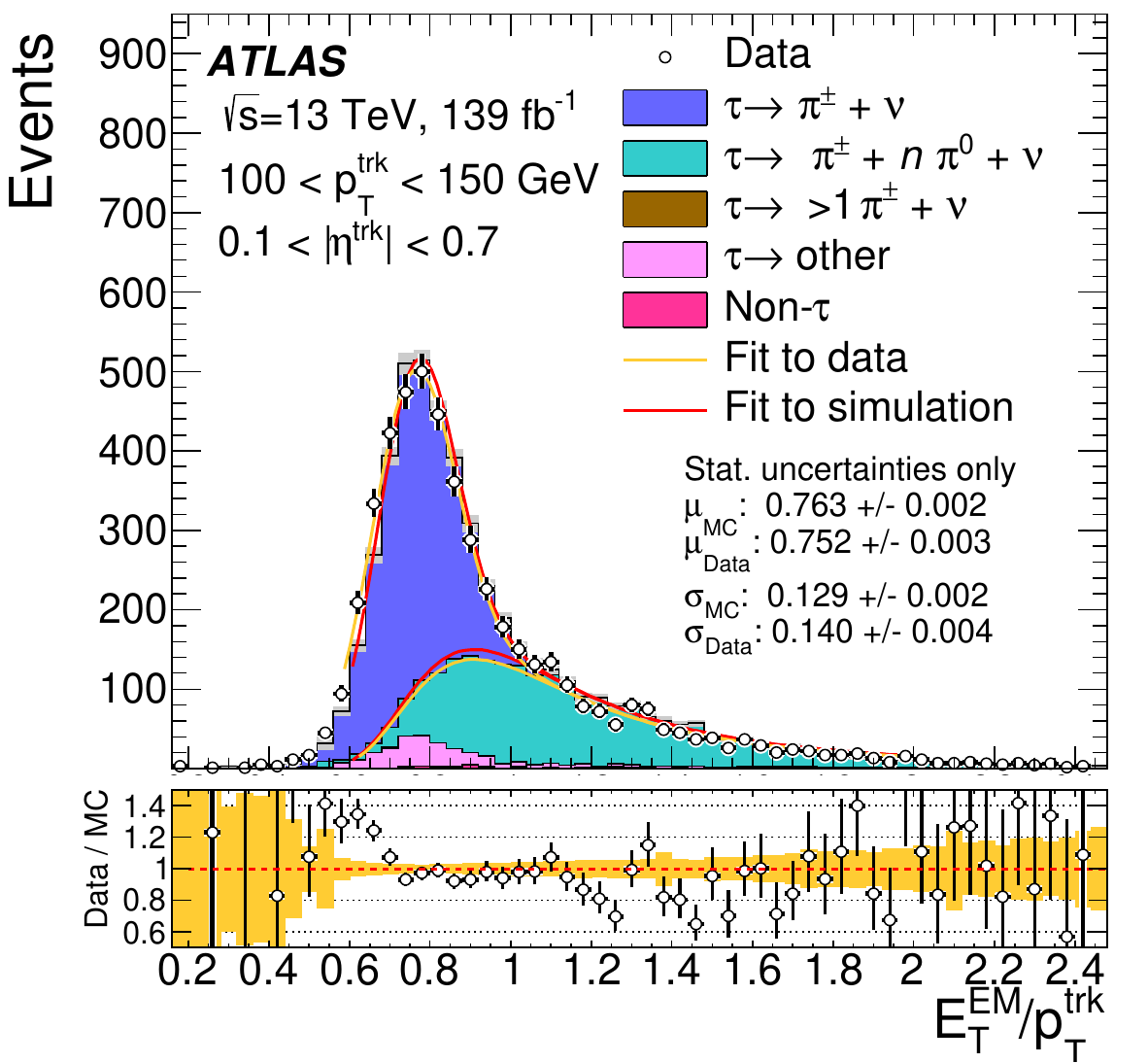}
\includegraphics[width=0.43\textwidth]{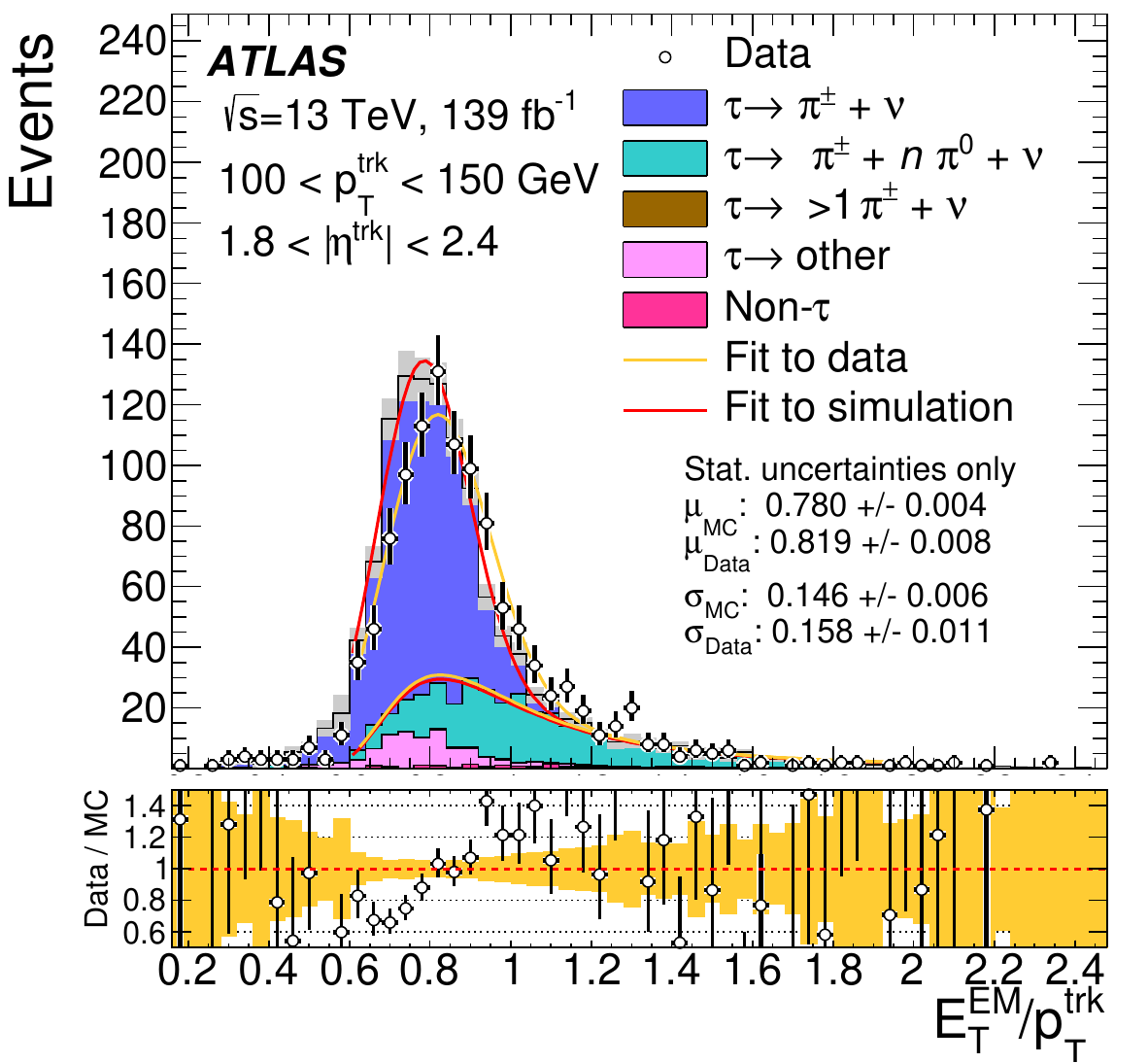}
\caption{Examples of the \EToverPT distribution and fit performed in the $16<\pTtrack<20$~\GeV\ (top), $30<\pTtrack<50$~\GeV\ (middle) and $100<\pTtrack<150$~\GeV\ (bottom) bins in the barrel (left) and endcaps (right). The data are shown by points with a stacked histogram for simulation. The component where $\tau$-leptons decay to multiple charged pions is small and difficult to see. The red fits correspond to simulation and the yellow fits correspond to data. Both the fits to the full distribution of a Gaussian+Landau function and just the Landau component from this fit are shown. The uncertainties shown are those from the statistics of the dataset and the limited Monte Carlo sample size. The bottom panel shows the ratio of data to simulation with statistical error bars on the data points and a yellow band illustrating the statistical uncertainty of the simulation. The simulation is normalised to the integral of the data.}
\label{fig.fitExample}
\end{figure}
 
The primary background is from events with $\tau$-leptons which decay to a charged pion, $\pi^\pm$, and at least one neutral pion, $\pi^0$. In the calorimeter, more energy will be reconstructed from this process for the same track momentum due to the electromagnetic shower from the $\pi^0$, which predominantly decays as $\pi^0\rightarrow\gamma\gamma$. This results in an upper tail in the \EToverPT distribution. Other single-charged-particle $\tau$-lepton decays, such as $\tau^\pm\rightarrow K^\pm\nu_\tau$, also contribute at a lower level.
 
The aim of this analysis is to extract the mean and width of a Gaussian function fitted to the $W^\pm\rightarrow\tau^\pm(\rightarrow\pi^\pm\nu_\tau)\nu_\tau$ signal in the core of the \EToverPT distribution to probe the calorimeter response and resolution. To extract this, the distribution is fit with the sum of a Gaussian function and another function that describes the background from all processes other than $W^\pm\rightarrow\tau^\pm(\rightarrow\pi^\pm\nu_\tau)\nu_\tau$. The functional form of this second term is taken to be a \textit{Landau} function~\cite{Landau:216256}. This is chosen empirically as it is seen to accurately capture the shape of the background. The range of the fit is then taken as the mean, $\mu_{\pi^\pm}$, of a Gaussian function describing \EToverPT for signal in simulation minus 1.5 times its width, $\sigma_{\pi^\pm}$, to 2, i.e.\ $[(\mu_{\pi^\pm}-1.5\sigma_{\pi^\pm}),2]$. The combined Gaussian+Landau function is fit to the distribution in data and simulation. In both cases the shape of the Landau function is fixed from a fit to the background processes in simulation. The three parameters of the Gaussian signal and the normalisation of the Landau function are allowed to float in the final fit. The mean of the Gaussian function gives the calorimeter response scale and the width of the Gaussian function measures the resolution of \EToverPT, which contains components from both the calorimeter and track reconstruction. Figure~\ref{fig.fitExample} shows the functions fitted to data and simulation in six of the (\pTtrack, $|\eta^{\textrm{trk}}|$) bins, and the extracted mean and width of the Gaussian function are displayed.
% End of text imported from the .//sections/Analysis.tex input file
 
\section{Calorimeter response measurement and comparison with simulation}
\label{sec:results}
% The next lines are included from the .//sections/Results.tex input file
The data are binned in both the track \pT and track $|\eta|$ to probe the calorimeter response at different momenta and in different regions. The calorimeter response is expected to vary significantly across $|\eta|$ due to the geometry and different calorimeter technologies. In the central region there is a large region which is expected to have a uniform response due to similar calorimeter structures up to $|\eta|=0.7$. As the rejection of muons is more difficult at $|\eta|\sim 0$ a \textit{barrel} selection region is defined by $0.1<|\eta|<0.7$. In this region there is a high-granularity liquid-argon/lead electromagnetic calorimeter and a steel/scintillator hadronic calorimeter. Another uniform detector region exists in the \textit{endcap} region of $1.8<|\eta|<2.4$, where there is liquid-argon/lead electromagnetic calorimetry and liquid-argon/copper hadronic calorimetry. Due to differences in technology and material in the two regions, differences in the modelling might be expected. To study the response across the detector, results are shown for three \pTtrack bins: $30<\pTtrack<50$~\GeV, $50<\pTtrack<70$~\GeV\ and $70<\pTtrack<100$~\GeV, as a function of $|\eta^{\textrm{trk}}|$.
 
Figure~\ref{fig.scaleVsPt} shows the fitted response for the \textit{barrel} and \textit{endcap} regions defined above, as a function of the track \pT. Clear differences are seen between the data and simulation in both regions, with a ${\sim}2\%$ overestimate of the response in simulation in the barrel and a ${\sim}4\%$ underestimate of the response in simulation in the endcaps relative to the data. The data/MC differences also follow the same trends as seen in the jet response in these two regions~\cite{JETM-2018-05}. It is checked that the fits give a reasonable $\chi^2$ per degree of freedom and are stable when the fitting range and binning of the data are altered. These plots only show the uncertainties from the limited number of events in the samples, and systematic uncertainties are discussed in the next section. At the highest energies the steeply falling momentum spectrum and the increasing track resolution produces a bias in the measured track momentum compared to the true pion transverse momentum, $p_\textrm{T}^\textrm{true}$, resulting in the distribution flattening in both data and simulation. The biases in simulation from $\langle p_\textrm{T}^\textrm{trk}/p_\textrm{T}^\textrm{true}\rangle$ in the central (endcap) region for the highest three \pTtrack bins are [0.2\%, 0.8\%, 1.7\%] ([1.7\%, 4.7\%, 10.8\%]).
 
Figure~\ref{fig.scaleVsEta} shows the fitted response across the detector in fine $|\eta^{\textrm{trk}}|$ bins for the three different \pTtrack bins which have the most events. The calorimeter structure is clearly seen and the simulation follows the data, but some differences are present. The largest disagreements are ${\sim}5\%$ in the $1.0<|\eta^{\textrm{trk}}|<1.2$ region. When the two sides of the detector are analysed independently the same trends are seen in the response. Additionally, when the analysis is performed separately for positively and negatively charged tracks consistent results are seen. In this energy regime the responses to $\pi^+$ and $\pi^-$ are expected to be very similar and this check also tests the alignment of the inner detector as alignment errors would have different effects on positive and negative tracks.
 
\begin{figure}
\centering
\includegraphics[width=0.45\textwidth]{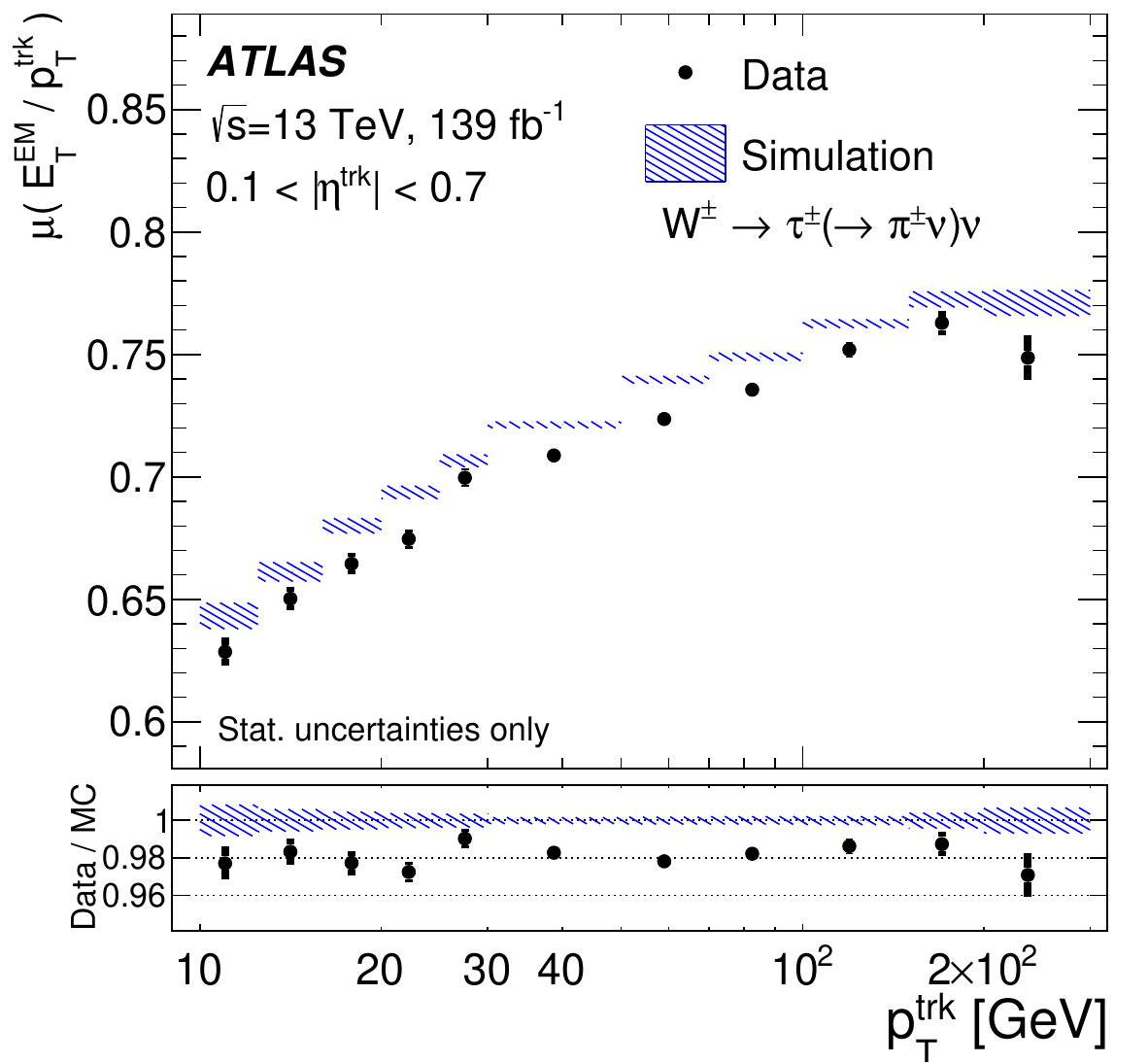}
\includegraphics[width=0.45\textwidth]{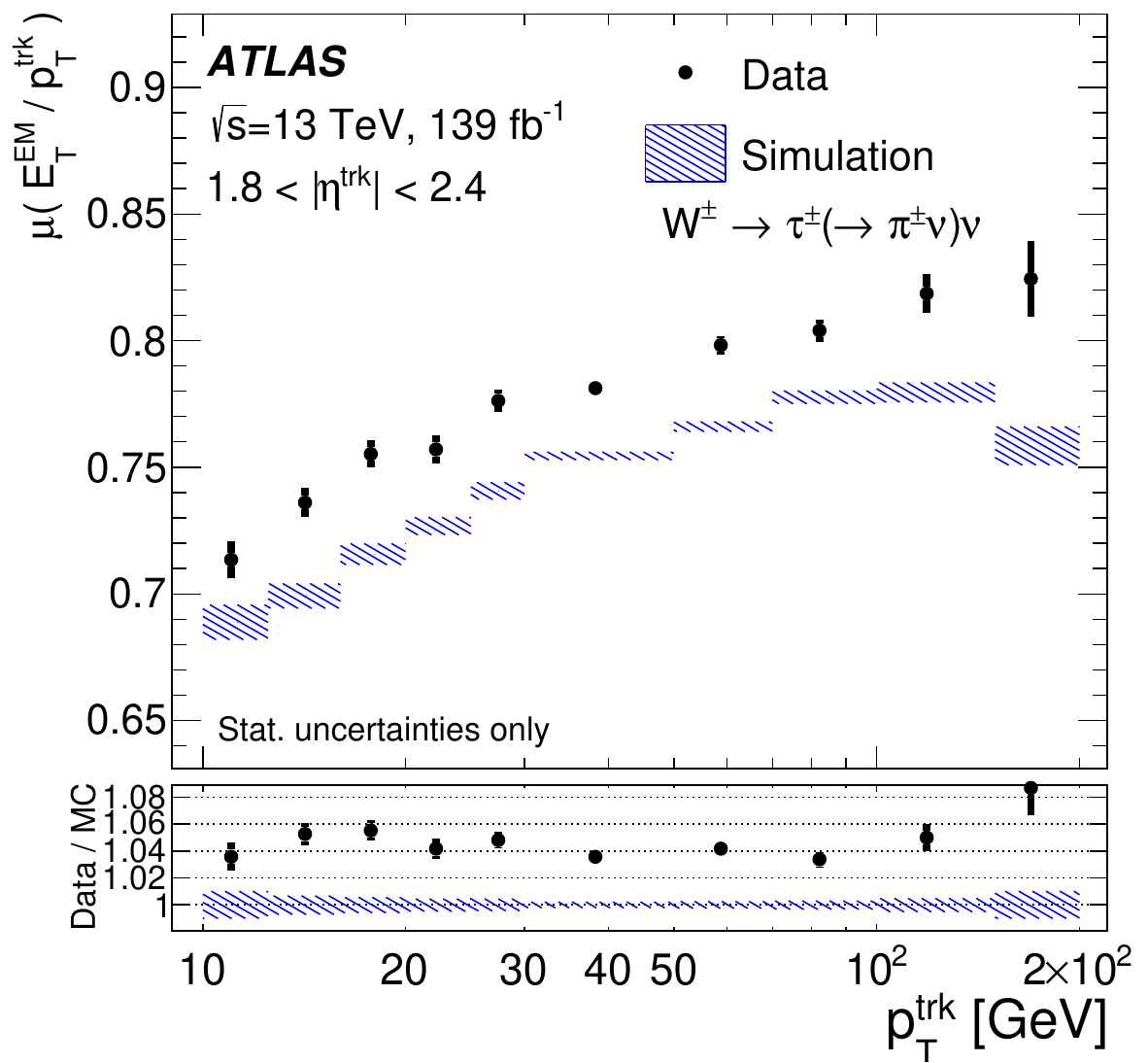}
\caption{The fitted mean of the signal \EToverPT distribution as a function of the track \pT in the central region (left) and endcap region (right). The error bars and uncertainty band show only the statistical uncertainties from the size of the dataset and simulated samples respectively.}
\label{fig.scaleVsPt}
\end{figure}
 
\begin{figure}
\centering
\includegraphics[width=0.32\textwidth]{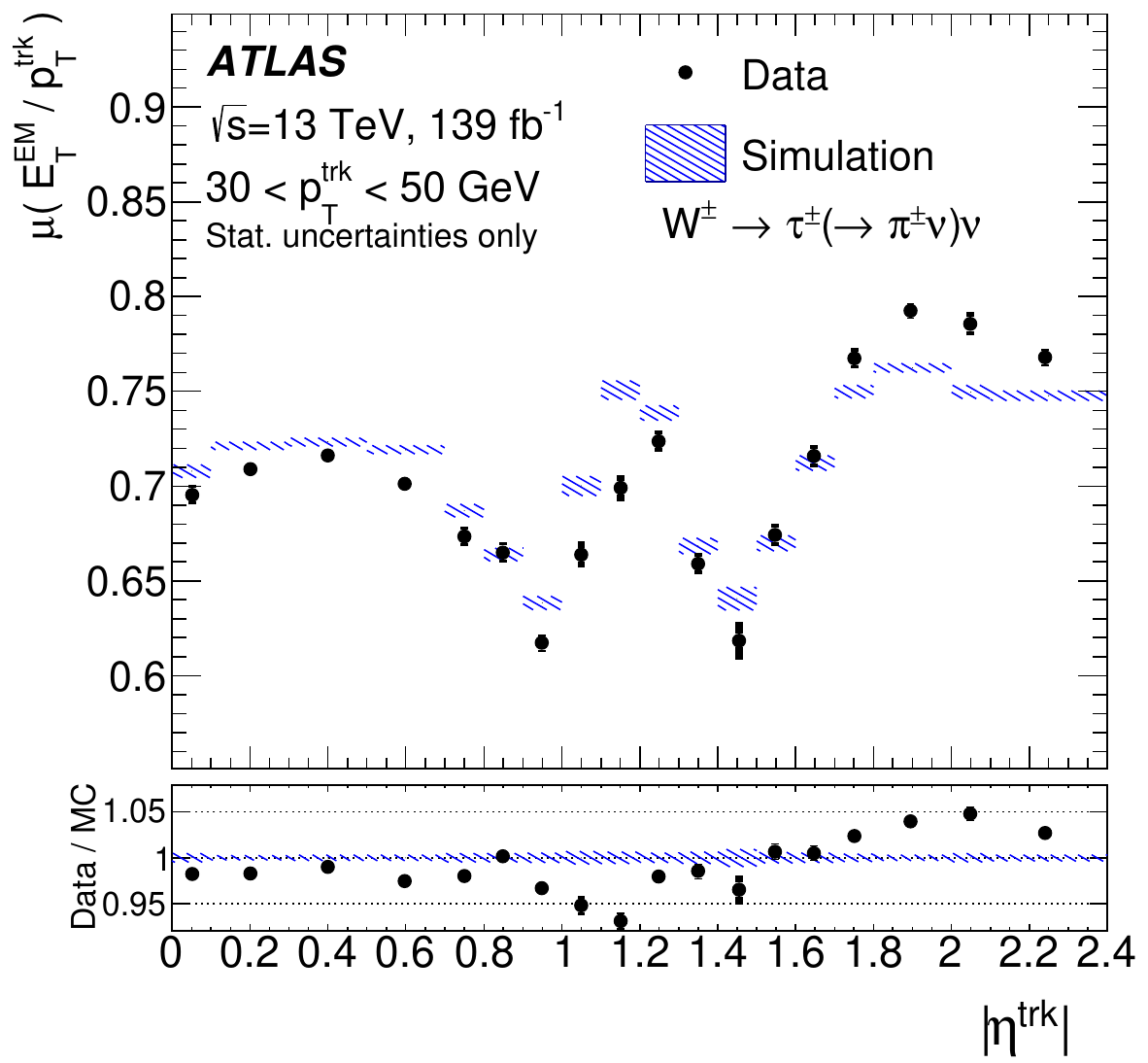}
\includegraphics[width=0.32\textwidth]{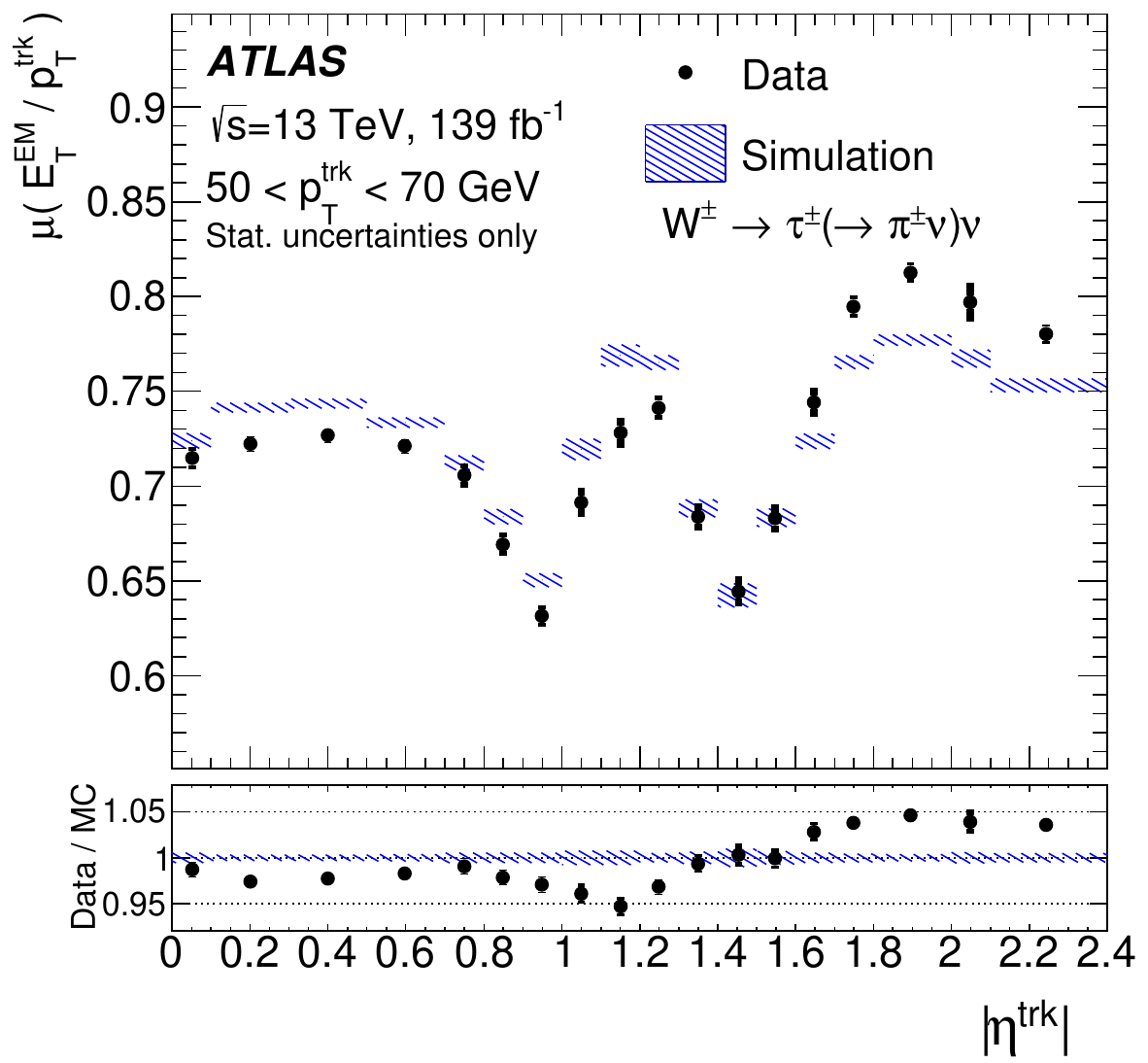}
\includegraphics[width=0.32\textwidth]{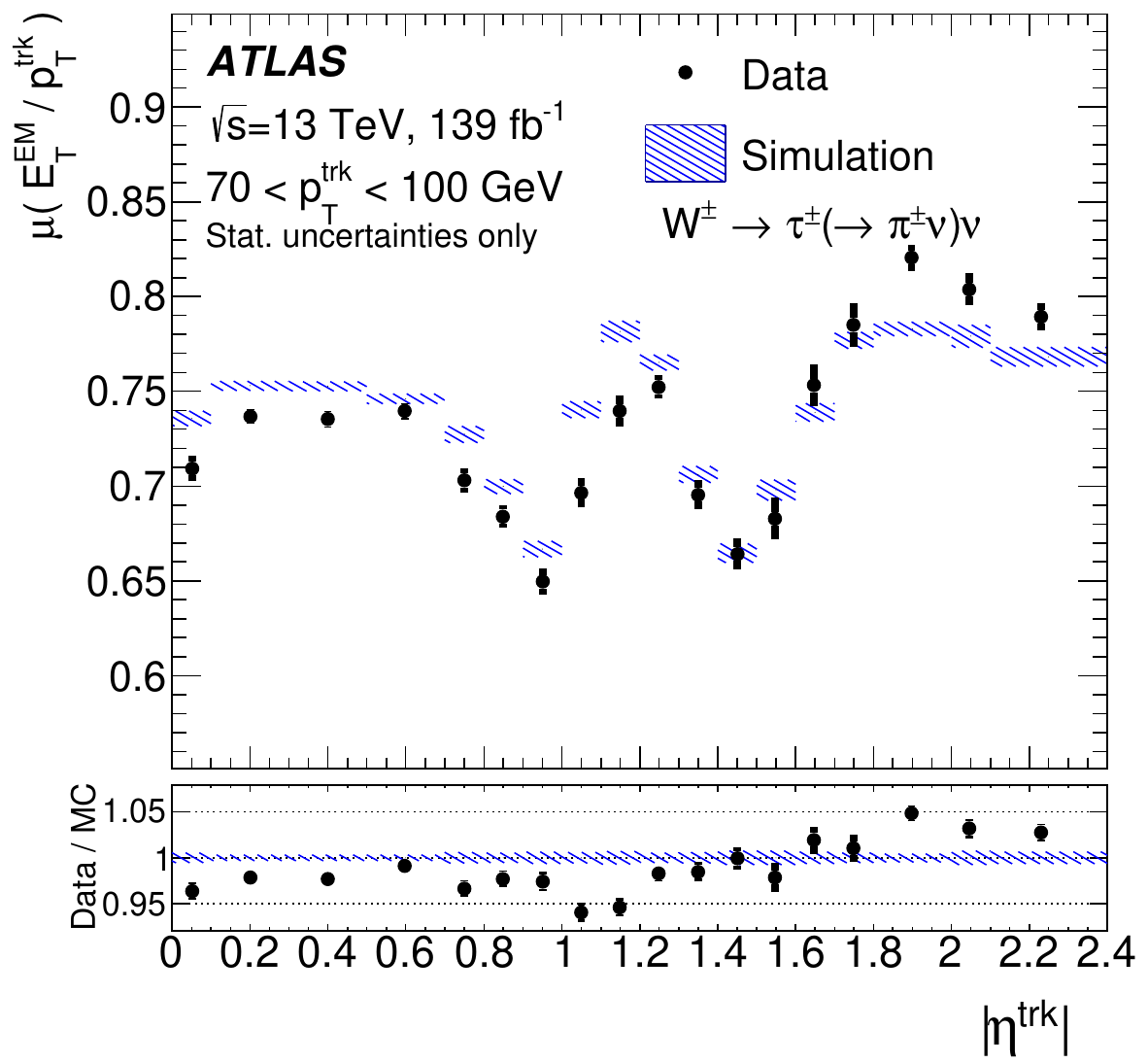}
\caption{The fitted mean of the signal \EToverPT distribution as a function of the track $|\eta^{\textrm{trk}}|$ in three different \pTtrack bins. The error bars and uncertainty band show only the statistical uncertainties from the size of the dataset and simulated samples respectively.}
\label{fig.scaleVsEta}
\end{figure}
 
% End of text imported from the .//sections/Results.tex input file
 
\section{Uncertainties in the response measurement}
\label{sec:systematics}
% The next lines are included from the .//sections/Systematics.tex input file
Several different effects could bias the measurements of the calorimeter response scale. These include uncertainties in the methodology of the fitting procedure, the backgrounds entering the event selection, the corrections for the effects of \pu, and the modelling of tracks. The methods used to assess each of the different sources are described below and are then summarised in Figures~\ref{fig.systematics} and~\ref{fig.systematicsEta}. All uncertainties which are determined from a difference with respect to the nominal result are symmetrised.
 
\paragraph{Fit function (closure):}A Gaussian+Landau fit to the signal and background is used to extract the calorimeter response scale. This is quantified in simulation by the difference between the fitted mean obtained from a Gaussian fit to just the signal $W^\pm\rightarrow\tau^\pm(\rightarrow\pi^\pm\nu_\tau)\nu_\tau$ events and the fitted mean of the Gaussian function obtained from a combined Gaussian+Landau fit to the simulated signal and background. This closure test is seen to perform very well, showing that the Landau function choice is appropriate. The difference between the two values obtained is taken as a systematic uncertainty of the measurement. It is found to be subdominant and typically at the ${\sim}0.2\%$ level.
 
\paragraph*{Bias in the fitted background shape:}The shape of the background is taken from a fit to simulation, which could be biased relative to the shape in data. The background is primarily $W^\pm\rightarrow\tau^\pm(\rightarrow\pi^\pm n\pi^0\nu_\tau)\nu_\tau$ events. Imperfections in the simulation of the shape of this background can come from detector effects such as the energy response to the charged and neutral pions, or from the energy spectra of the charged and neutral pions as modelled by the MC generator. The energy scale and resolution for electromagnetic showers, such as those from $\pi^0\rightarrow\gamma\gamma$ decays, has been measured in ATLAS~\cite{EGAM-2018-01} and is known better than that for the hadronic showers probed here. The impact of detector mismodelling is tested by shifting or smearing the simulated response and evaluating the change in the fit when it is applied to data with the altered background shape. The hadronic response is shifted down (up) by 3\% (5\%) in the central (endcap) region, and separately smeared to increase the resolution by 10\%. These values are determined from the discrepancies between data and simulation observed in this analysis. The smearing in the central region, and scale variation in the endcaps, produce significant uncertainties in the final results. To assess the modelling of the shape an alternative generator is used for the $W+$jets background; no significant impact on the shape is seen, so this is neglected. Therefore, the uncertainty in the shape of the fitted background is dominated by the detector modelling.
 
\paragraph*{Electron and muon contamination:}Contamination from electrons and muons is expected to be small: less than 0.4\% and 0.7\% respectively. The residual contamination is tested by tightening the criteria to reject these events: requiring the fraction of energy in the hadronic calorimeters to be ${>}20\%$, and that there are no associated muon segments. Changes in the measured response in data define the uncertainty for muons. For electrons the difference between the changes in the measured response in data and simulation defines the uncertainty because the hadronic energy fraction cut can also bias the signal. Both of these uncertainties are found to be smaller than the statistical uncertainties of the measurement.
 
\paragraph*{Non-lepton backgrounds:}There is a small amount of background where the track does not originate from a lepton. The $Z(\rightarrow\nu\nu)+$jets contamination can be reduced by a tighter selection on the transverse mass of the track and the \met{}. The deviation from the nominal response when the upper bound on $m_\textrm{T}$ is reduced by 10~\GeV\ is taken as an uncertainty in this background. Additionally, the deviation from the nominal response when events with $m_\textrm{T}<10$~\GeV\ are excluded is taken as an uncertainty in the potential contamination from events where the \met{} comes from detector effects. These are also found to be smaller than the statistical uncertainties.
 
\paragraph*{\Pu uncertainties:}Two uncertainties in the subtraction of the contributions from \pu are considered. Non-linearities in the required corrections can lead to imperfections in the removal of the effects of \pu. The \textit{closure} of the pile-up corrections is tested by taking half the difference between the responses for two large single-particle samples with different \pu, $\langle\mu^{\textrm{PU}}\rangle\sim20$ and $\langle\mu^{\textrm{PU}}\rangle\sim50$, after the corrections. Additionally, the \textit{pile-up residual} correction is determined from simulation where \pu might not be well modelled. Studies of the $\mu^{\textrm{PU}}$ dependence of the median \pT density of the event, $\rho$, the number of clusters, and the cell energy and noise find that the modelling of the energy flow in the simulation is within 10\%--20\% of the data across the detector. Therefore, 25\% of the residual correction is taken as an uncertainty in the modelling of \pu, which forms a minor uncertainty in the analysis. In data it is checked that the fitted response is stable with respect to \pu by splitting the dataset into two independent datasets via high \pu and low \pu selections based on $\mu^{\textrm{PU}}$, and also into three independent datasets, and no significant differences given the statistical uncertainties are seen in the measured response. 
 
\paragraph*{Tracking uncertainties:}Any bias in the momentum scale of the reconstructed tracks directly affects the measurement. Muons are precisely calibrated using $Z\rightarrow\mu\mu$ and $J/\psi\rightarrow\mu\mu$ events. The analysis selection is modified by removing the $m_\textrm{T}$ and transverse impact parameter requirements and requiring that a reconstructed muon is matched to the isolated track. The \pT of the track and muon are obviously correlated but fits to $p_\textrm{T}^\textrm{trk}/p_\textrm{T}^\mu$ probe the inner-detector track scale relative to the calibrated muons. Two components are considered, the uncertainty in the calibration of muons and the statistical uncertainties of this cross-calibration, as none of the deviations of simulation from data are significant. Additionally, differences in the track resolution can bias the measurement because of the underlying falling \pTtrack spectrum. The uncertainty in the momentum resolution of tracks is determined using the muons as described in Ref.~\cite{PERF-2014-05}. Tracks are smeared to increase the resolution in simulation within the uncertainties and the impact on the parameter of interest is symmetrised and taken as an uncertainty.
 
The various systematic uncertainties are smoothed~\cite{friedman1974data} across \pT as they are expected to vary smoothly. Across $|\eta|$, changes in the detector technology could result in sharp changes in the size of the systematic uncertainties. However, no such variations are seen, so the systematic uncertainties are also smoothed in the measurements as a function of $|\eta|$. The statistical components in data and simulation are not smoothed. The statistical uncertainty of the data is the largest uncertainty component in much of the phase space covered by the analysis.
 
Figure~\ref{fig.systematics} shows the uncertainty from each source as a function of track \pT for the central and  endcap regions. The uncertainty from imperfect knowledge of the background width (scale) in the central (endcap) region, and \pu uncertainties in the endcap region, are the only uncertainties that are larger than the statistical uncertainty of the dataset or the simulated event samples in certain bins. The \pu uncertainties have some importance at low \pT, and the tracking uncertainties grow at large track momentum. Overall, in the central region the total uncertainty is ${\lesssim}1\%$ for $20<\pT<200$~\GeV\ and reaches ${<}0.6\%$ in the most precise region. The test-beam uncertainties are about $2\%$ in the barrel region~\cite{Abat:2010zza}, demonstrating the power of this in situ technique. In the  endcaps the uncertainties are slightly larger but are ${\lesssim}1\%$ for $30<\pT<100$~\GeV. Figure~\ref{fig.systematicsEta} shows the uncertainties in the measurements as a function of $|\eta^{\textrm{trk}}|$. The uncertainties are slightly larger due to the limited number of events in the smaller bins used to make fine calorimeter structures visible, but are otherwise dominated by the same sources as in the measurements as a function of \pT.
 
\begin{figure}
\centering
\includegraphics[width=0.45\textwidth]{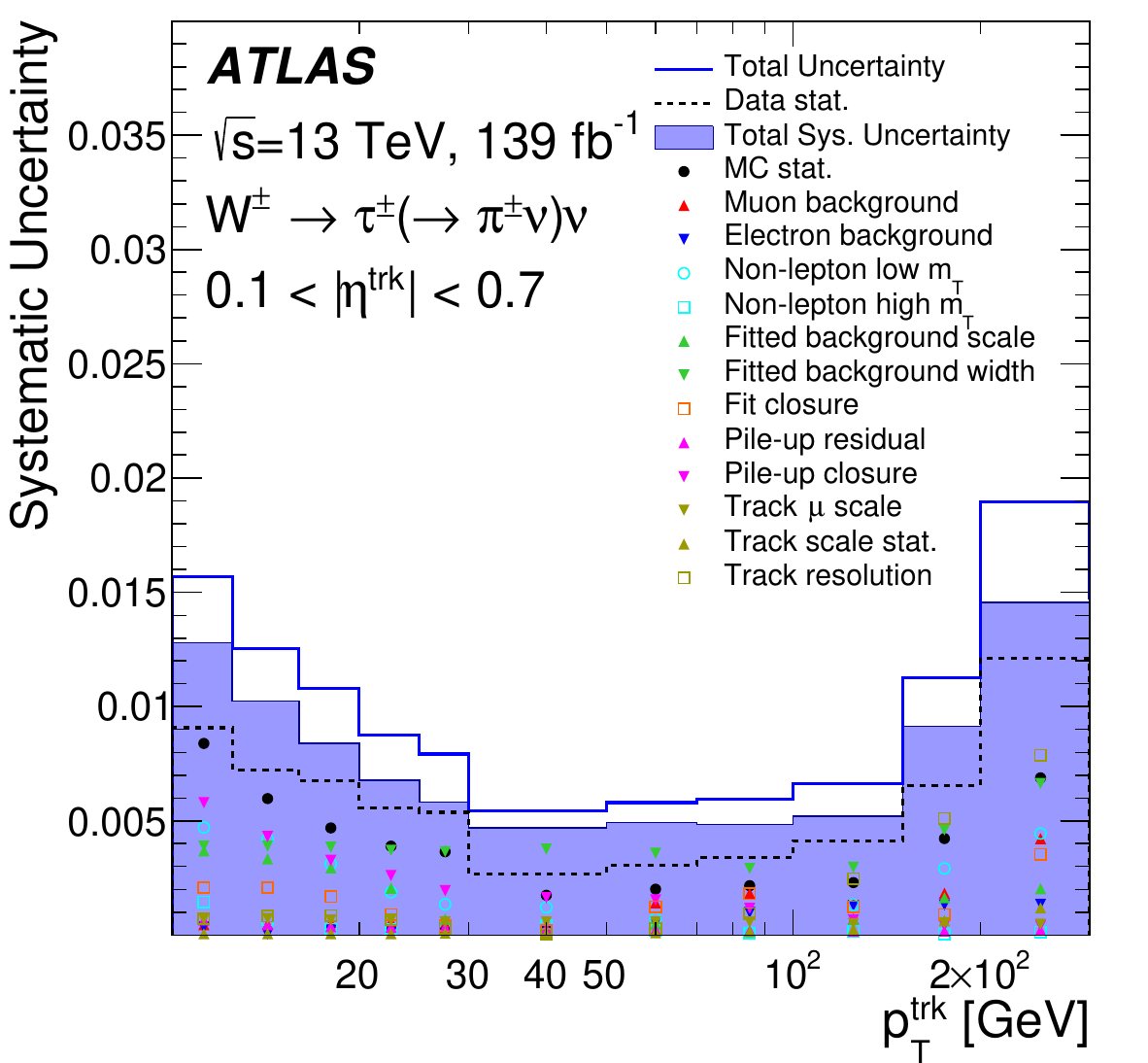}
\includegraphics[width=0.45\textwidth]{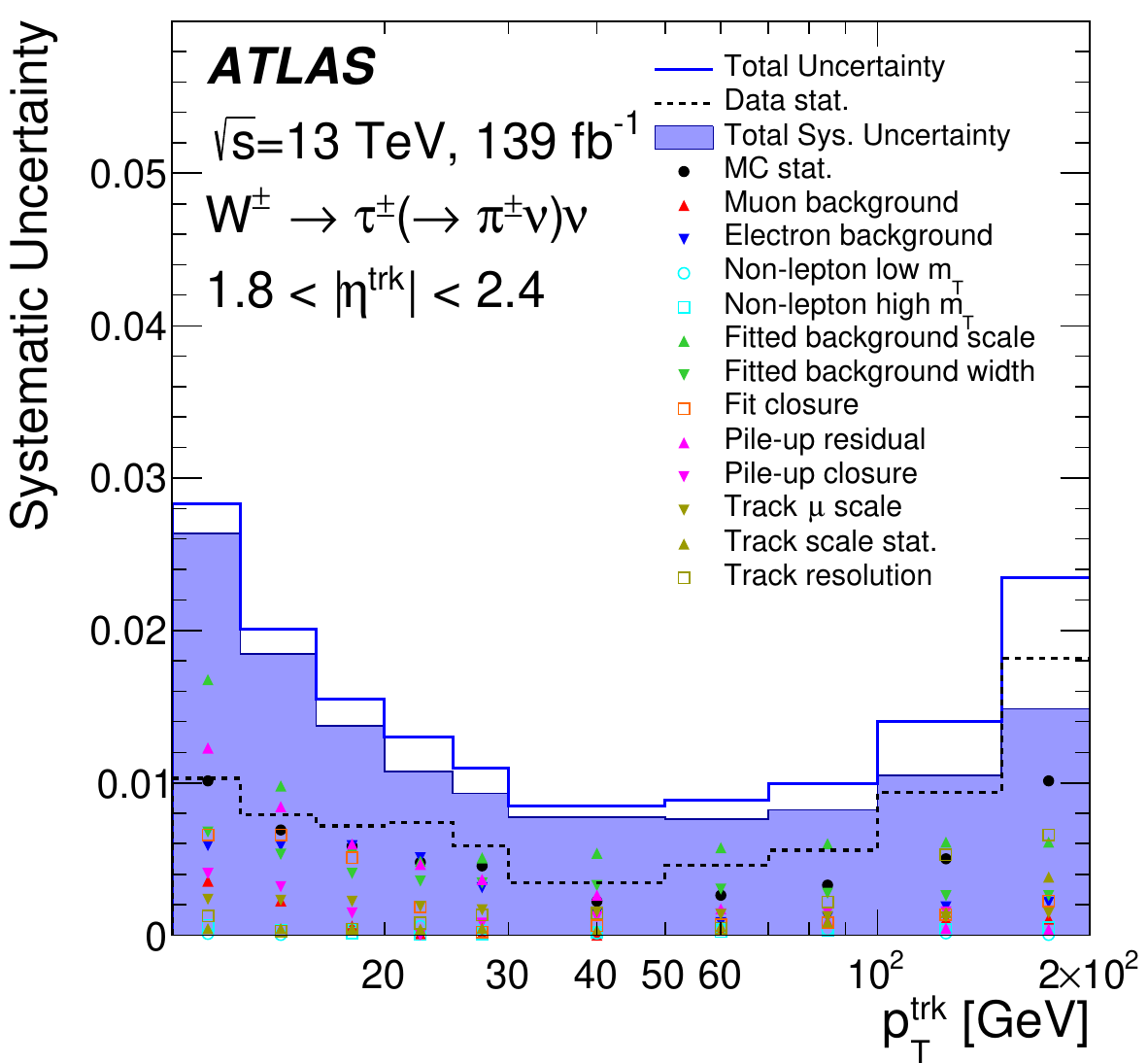}
\caption{The various systematic uncertainties which affect the measurement of the calorimeter energy scale for charged pions as a function of \pTtrack in the central (left) and  endcap (right) regions.}
\label{fig.systematics}
\end{figure}
 
\begin{figure}
\centering
\includegraphics[width=0.32\textwidth]{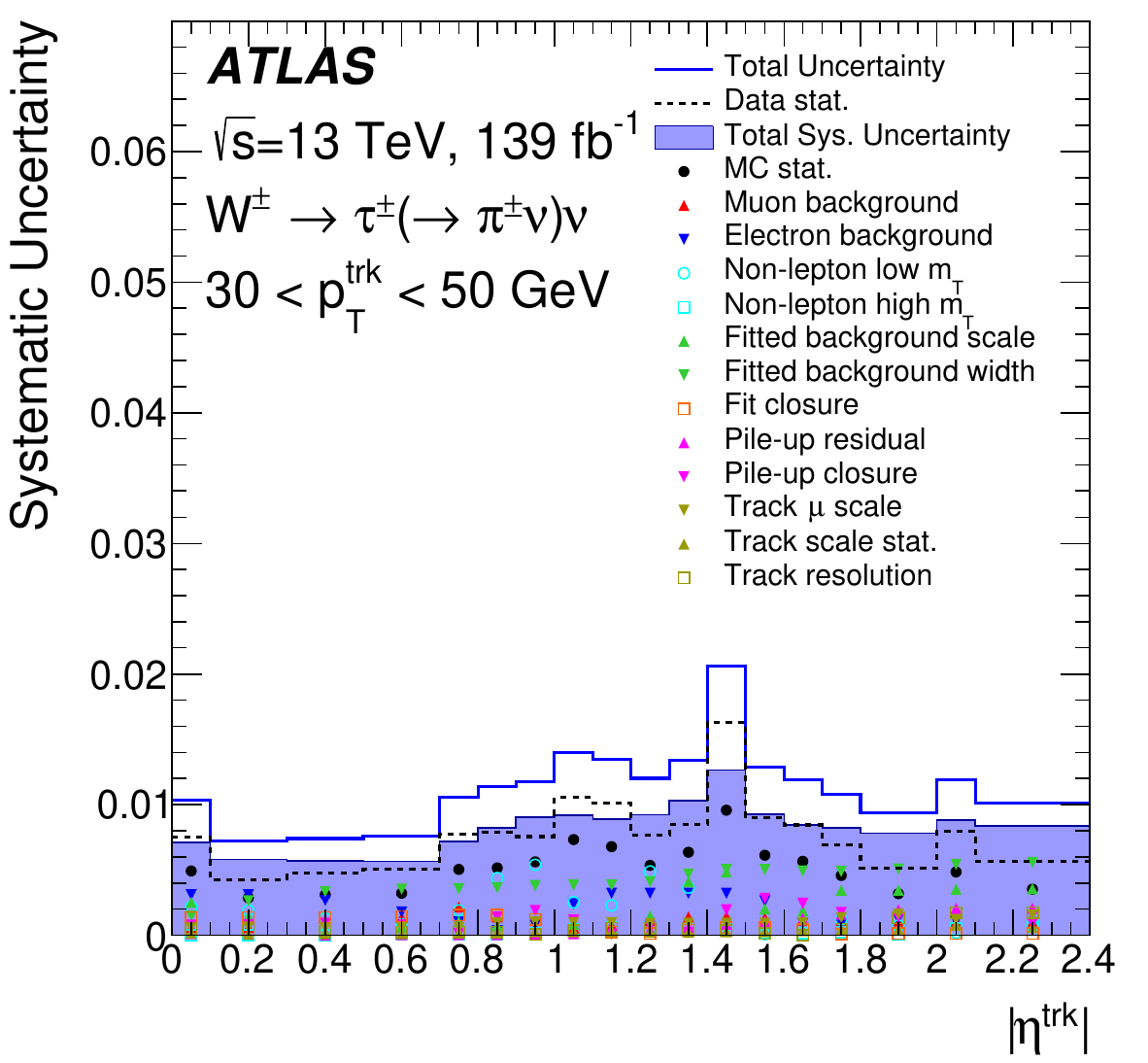}
\includegraphics[width=0.32\textwidth]{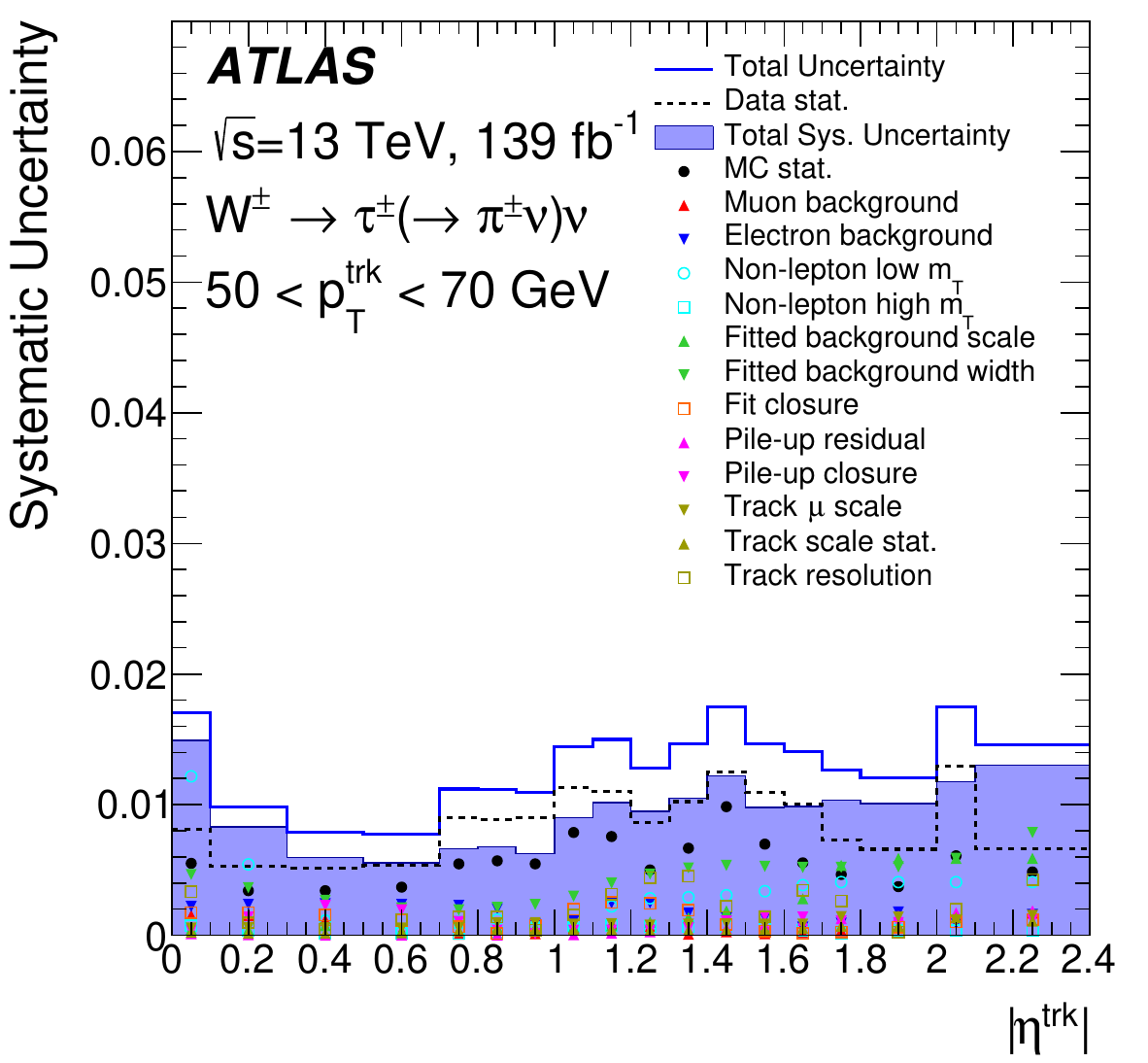}
\includegraphics[width=0.32\textwidth]{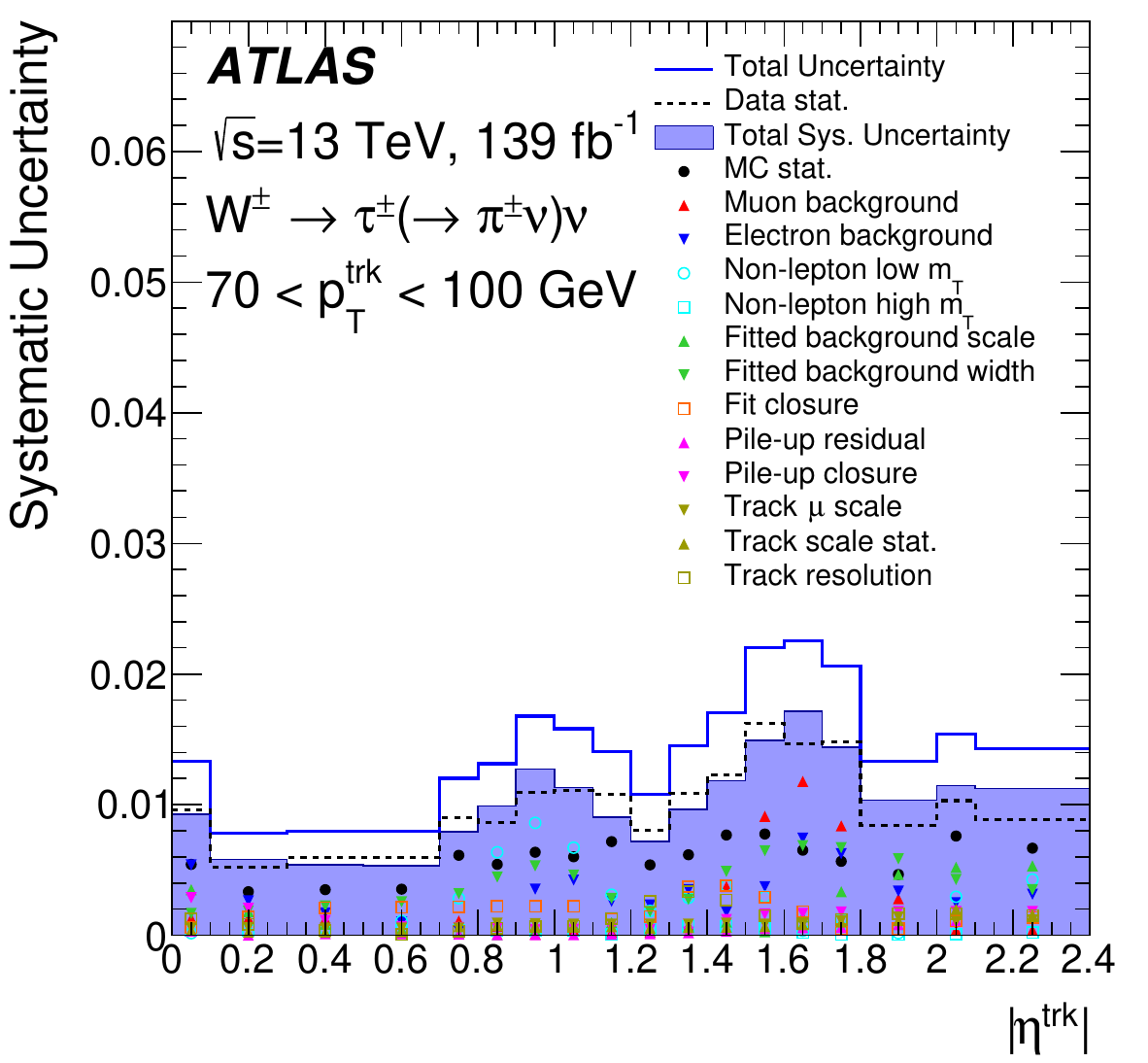}
\caption{The various systematic uncertainties which affect the measurement of the calorimeter energy scale for charged pions as a function of $|\eta^{\textrm{trk}}|$ for three \pTtrack ranges: $30<\pTtrack<50$~\GeV, $50<\pTtrack<70$~\GeV\ and $70<\pTtrack<100$~\GeV.}
\label{fig.systematicsEta}
\end{figure}
 
Having established the full uncertainties in the measurement, a calibration can be derived which can form an input to the high-\pT jet uncertainties~\cite{JETM-2018-05} and also to the  scale uncertainties for hadronic $\tau$-lepton decays~\cite{PERF-2013-06}. Figure~\ref{fig.calibration} shows the measured data to simulation ratio of the response and its uncertainty in different \pTtrack bins in the central and endcap regions. It is expected that such a calibration will be smooth across \pT. The same procedure as used for the jet energy scale is used~\cite{PERF-2012-01,JETM-2018-05} to translate the binned measurements and uncertainties into smooth uncertainty eigenvectors and a smooth calibration curve. Each of the measurements is divided into finer bins of 0.1~\GeV\ using second-order polynomial splines. The final calibration curve is determined by smoothing the measurements with a Gaussian kernel of varying width. Each individual uncertainty component is treated as correlated across \pT and is then propagated through the same procedure after varying it by 1$\sigma$. The difference between the calibration curve with the shifted systematic uncertainty input and the nominal calibration curve is taken as 1$\sigma$ in the varied uncertainty. Each uncertainty source is treated as correlated across \pT and uncorrelated with all other sources of uncertainty. This smoothing procedure reduces statistical fluctuations in the central values and in each propagated uncertainty component. Therefore, the final uncertainties are slightly smaller than those for the corresponding individual measurement points. The resulting smooth calibration curve and uncertainty band are shown in Figure~\ref{fig.calibration}.
 
Figure~\ref{fig.calibrationEta} shows the data/MC ratio as a function of $|\eta^{\textrm{trk}}|$ for the three measured \pT ranges along with the associated uncertainties. As the changes in detector technology can result in sharp features in the calibration as a function of $|\eta^{\textrm{trk}}|$ a smooth calibration curve is not derived for these results.

\begin{figure}
\centering
\includegraphics[width=0.45\textwidth]{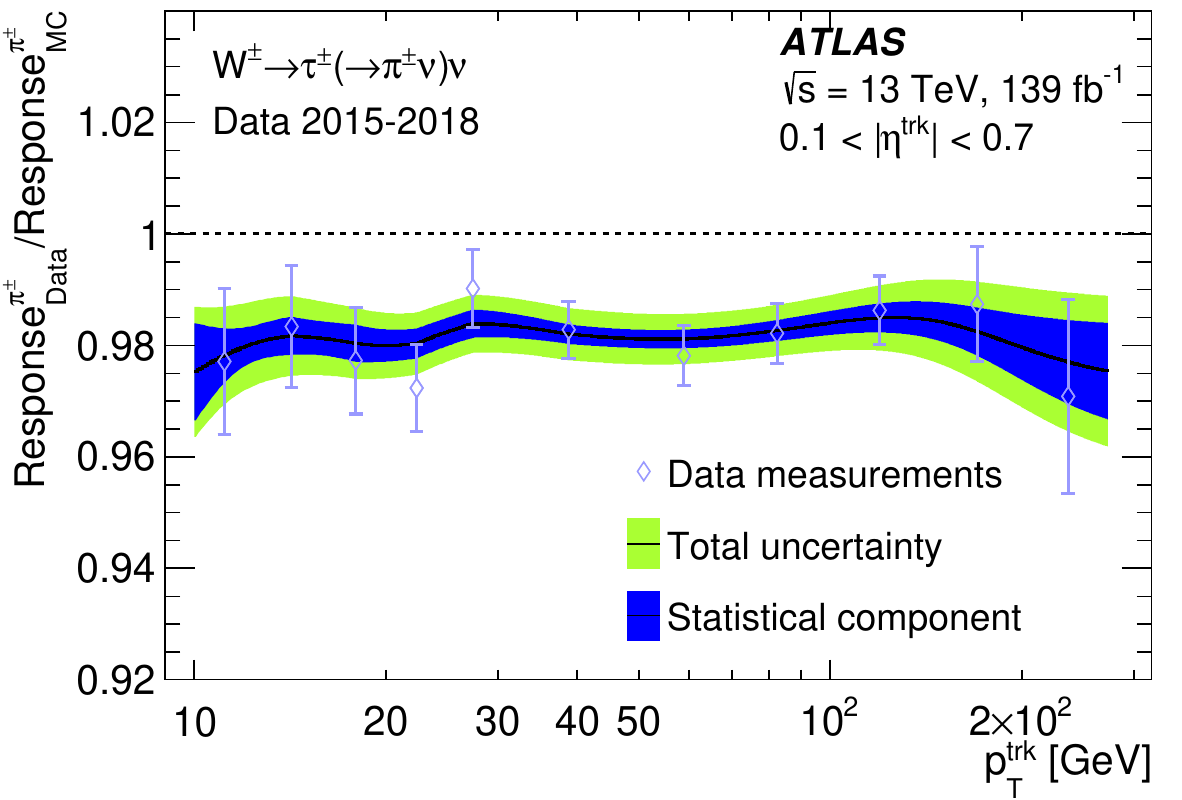}
\includegraphics[width=0.45\textwidth]{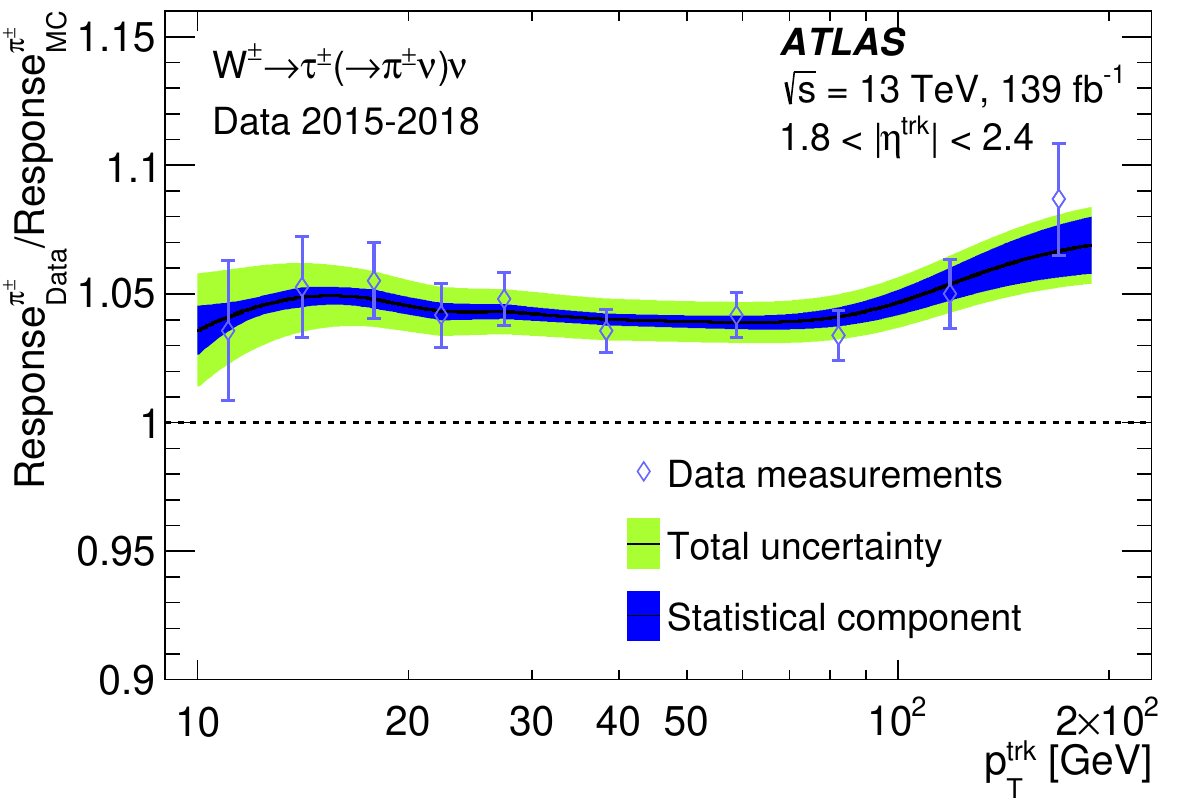}
\caption{The measured data to simulation ratio of the response as a function of \pTtrack for the barrel (left) and  endcaps (right). Points represent the different measurement bins with their total uncertainty from both the statistical and systematic sources. Also shown is the smooth calibration curve with its associated uncertainties.}
\label{fig.calibration}
\end{figure}
 
\begin{figure}
\centering
\includegraphics[width=0.32\textwidth]{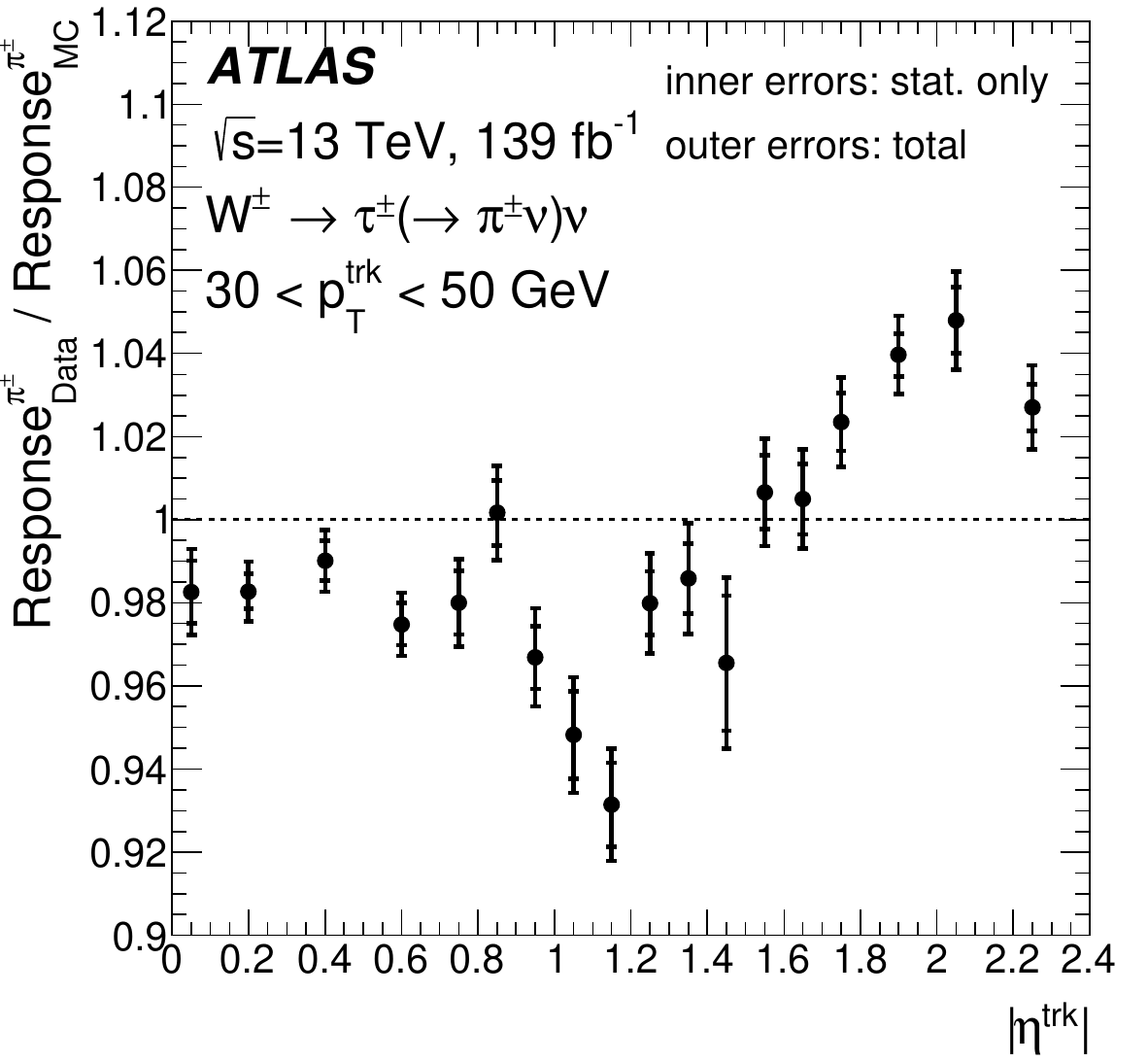}
\includegraphics[width=0.32\textwidth]{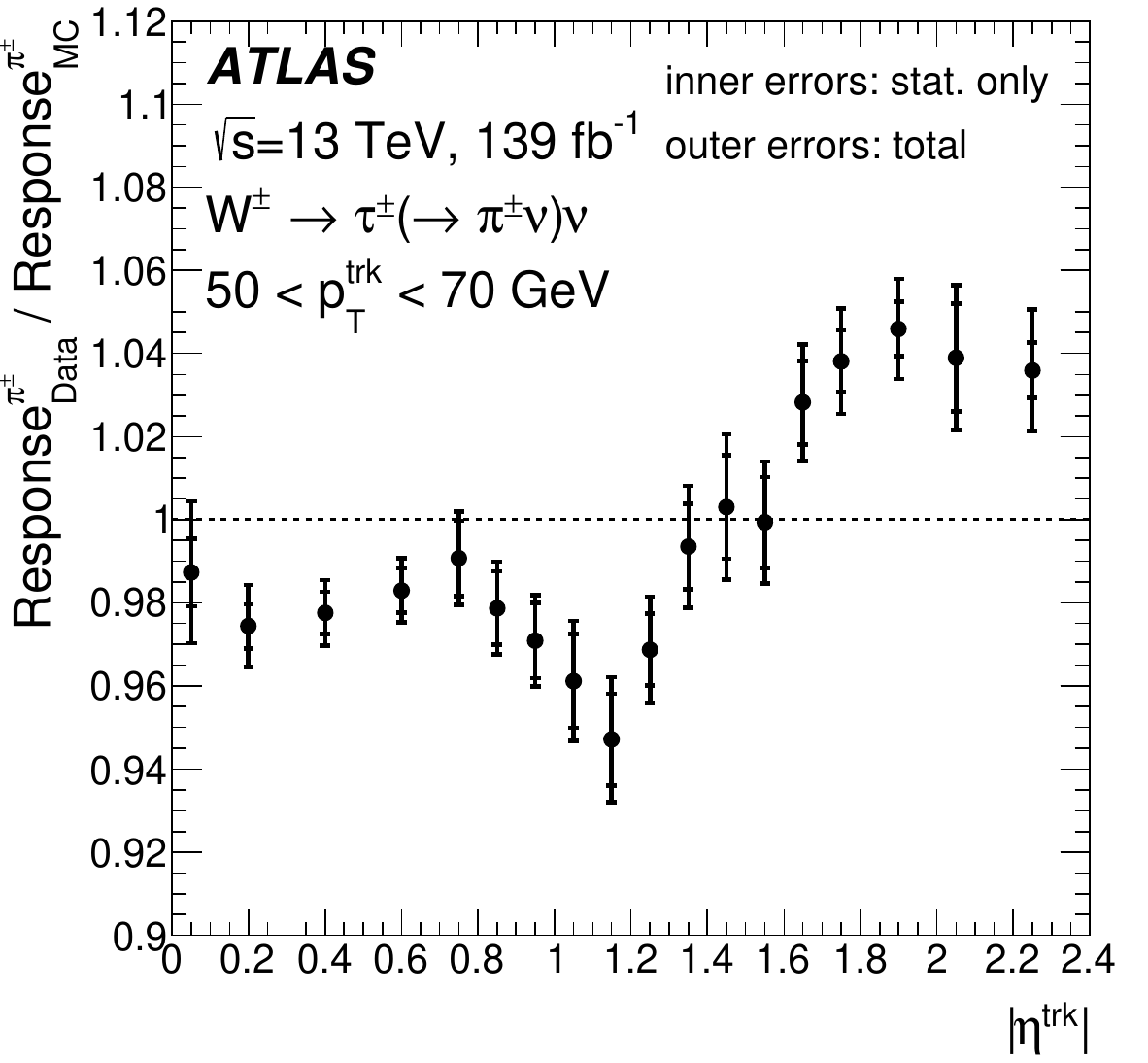}
\includegraphics[width=0.32\textwidth]{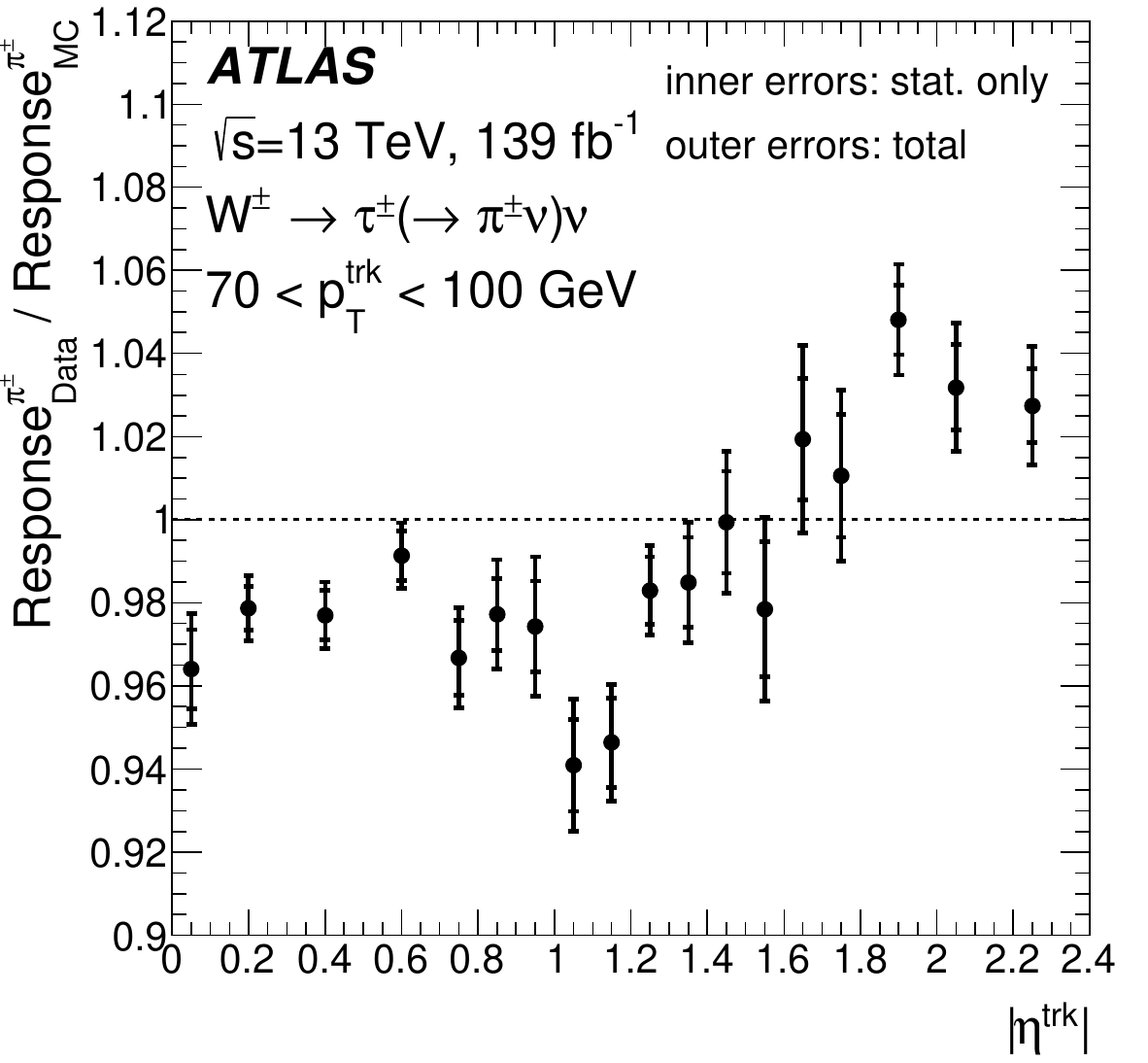}
\caption{The measured data to simulation ratio of the response as a function of $|\eta^{\textrm{trk}}|$ for the three \pTtrack ranges: $30<\pTtrack<50$~\GeV, $50<\pTtrack<70$~\GeV\ and $70<\pTtrack<100$~\GeV. Inner error bars represent uncertainties from limited sample size and the outer error bars give the total uncertainty including both the statistical errors and systematic uncertainties.}
\label{fig.calibrationEta}
\end{figure}
% End of text imported from the .//sections/Systematics.tex input file
 
\section{Width of the \EToverPT distribution}
\label{sec:resolution}
% The next lines are included from the .//sections/Width.tex input file
The width of the \EToverPT distribution probes the resolution of both the tracker and the calorimeter; however, the tracking component is relatively small outside of the highest energy bins. The width will also contain components from the modelling of \pu noise falling into the region of the calorimeter considered in the measurement. These can be reliably corrected for in measurements of the scale, but the effect on the resolution is much harder to mitigate, so the level of agreement between simulation and data, which both include these effects, is considered. The increase in the width of \EToverPT due to the effects of \pu is measured by looking at the difference between the width in two simulated samples with $\langle\mu^{\textrm{PU}}\rangle\sim20$ and $\langle\mu^{\textrm{PU}}\rangle\sim50$. For the central (endcap) region the difference is found to be 8\% (16\%) for $16<\pTtrack<20$~\GeV\ and 3\% (12\%) for $30<\pTtrack<150$~\GeV. The impact of mismodelling the \pu is expected to be small for most of the \pTtrack spectrum since the simulation includes \pu and should capture most of the effect. Figure~\ref{fig.resVsPt} shows the relative width of the \EToverPT distribution as a function of track \pT in the central and endcap regions. In the central region the simulation shows about $10\%$ better resolution than the data, while in the endcaps the level of agreement is generally better. The relative width across $|\eta|$ is shown for three \pTtrack bins in Figure~\ref{fig.resVsEta}. In both figures the resolution for tracks selected in simulation is found to contribute significantly at high momentum, but across much of the spectrum it is significantly smaller than the calorimeter contributions, indicating that the discrepancies between simulation and data are more likely due to the simulation of the calorimeter response. 
 
In the central region where the statistical uncertainties are lower, $30<\pTtrack<150$~\GeV, it is possible to check if any of the systematic effects considered in Section~\ref{sec:systematics} could be the source of the discrepancies. The closure of the fits to obtain the width of the signal component is very good, with differences only at the level of ${\lesssim}2\%$ at \pTtrack of 30~\GeV\ and ${\lesssim}1\%$ above 50~\GeV. Tightening the selections to reject electrons and muons has a minimal effect on the measured resolution, ${\lesssim}0.5\%$. Shifting the response scale of the background and smearing the background response are each seen to change the resolution in data by ${\lesssim}1\%$ in this region of phase space. Tightening the criteria to reject non-lepton backgrounds also results in small changes of ${\lesssim}0.5\%$ beyond 30~\GeV. Therefore, resolution discrepancies of ${\sim}10\%$ between data and simulation are significant in this \pTtrack range compared to the potential systematic effects on the extraction of the \EToverPT width. Below 30~\GeV\ the fit closure grows to up to 6\% and the uncertainties related to the fitted background shape and the non-lepton backgrounds all increase to ${\lesssim}1.5\%$, so that at low momentum the discrepancies are less significant.
 
This supports the observation in the jet resolution measurement using \pT balance in dijet events that the resolution in simulation is slightly superior to that in data~\cite{JETM-2018-05} at medium to high \pT and this is therefore an area where it is desirable to improve the simulation.

\begin{figure}
\centering
\includegraphics[width=0.45\textwidth]{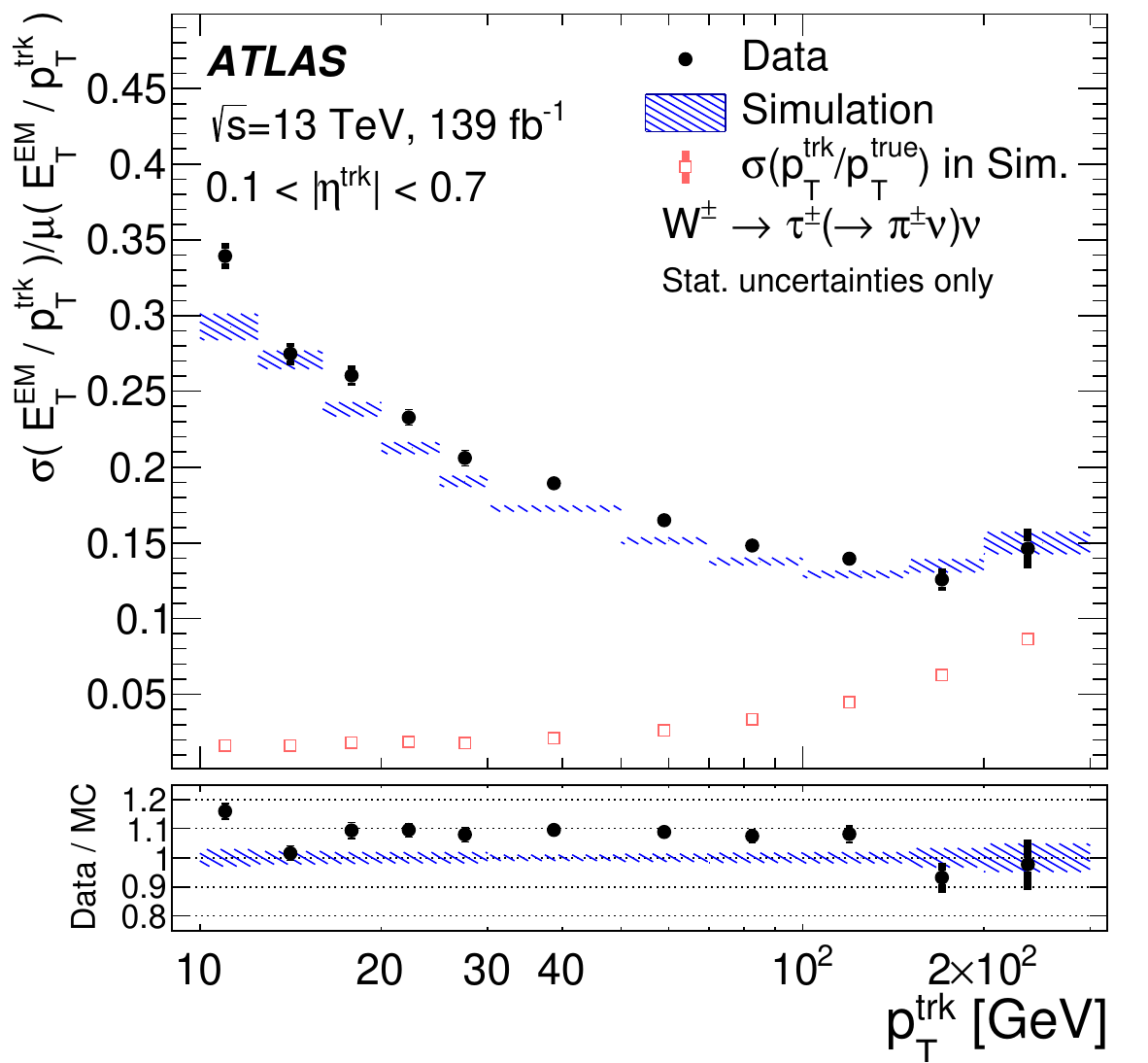}
\includegraphics[width=0.45\textwidth]{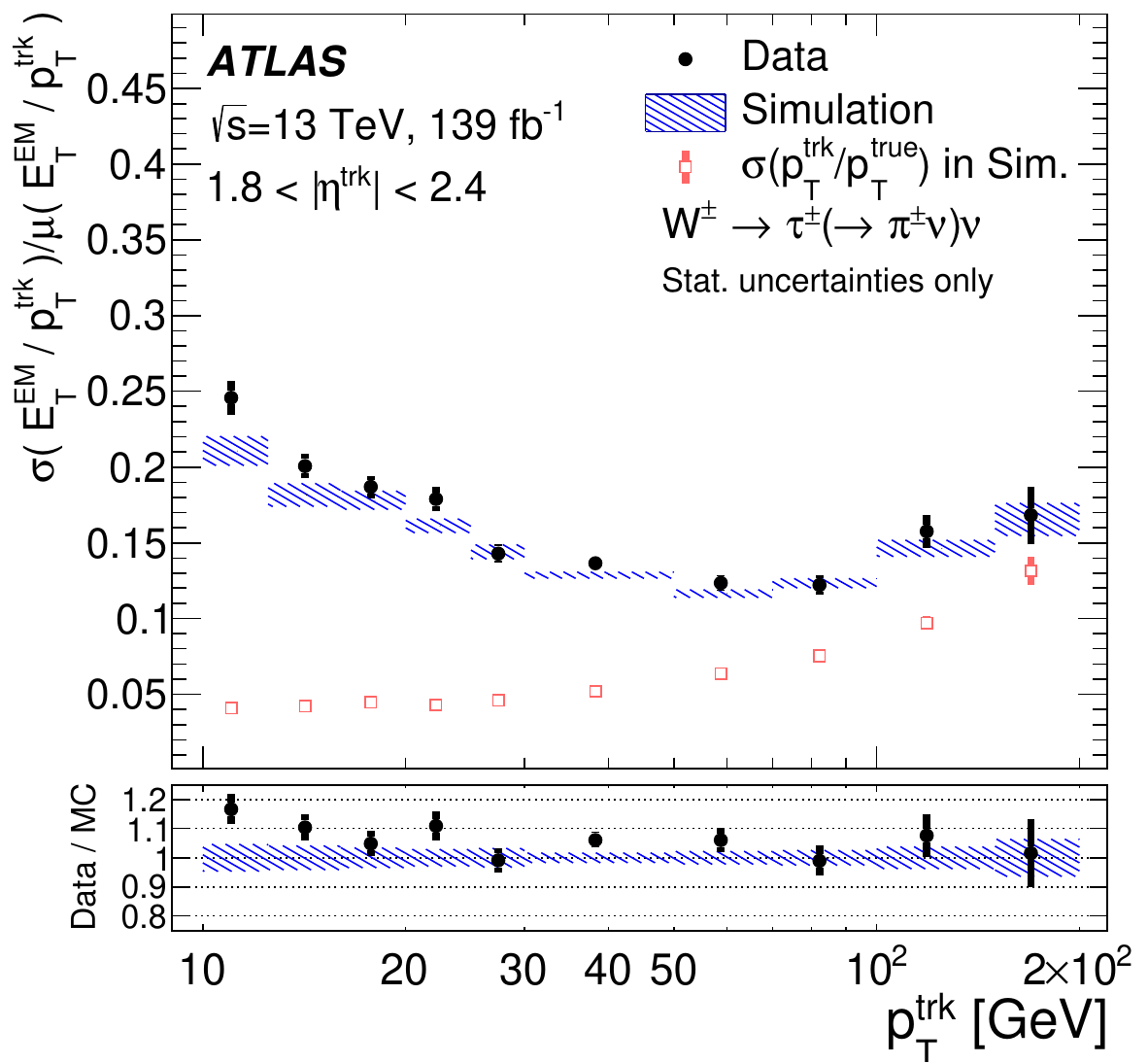}
\caption{The fitted width divided by the mean of the signal \EToverPT distribution as a function of the track \pT in the central region (left) and endcap region (right). Also shown is the fitted width of the track \pT divided by the generator-level \pT for true $\tau\rightarrow\pi^\pm\nu_\tau$ events in simulation to give an illustration of the contribution to the total width from the resolution of reconstructed tracks. The uncertainties shown are only those from the limited number of events in the dataset and simulated samples.}
\label{fig.resVsPt}
\end{figure}
 
\begin{figure}
\centering
\includegraphics[width=0.32\textwidth]{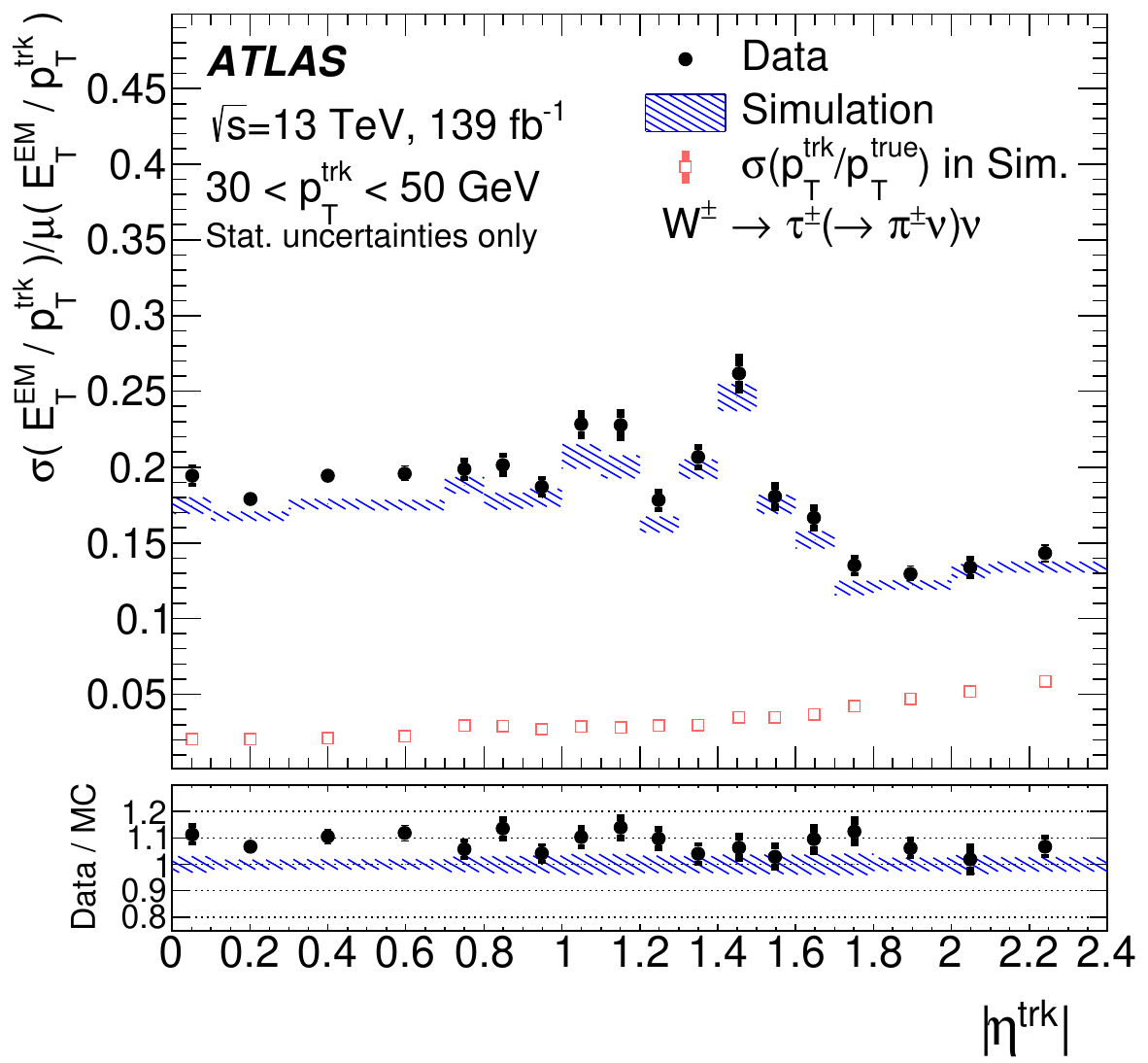}
\includegraphics[width=0.32\textwidth]{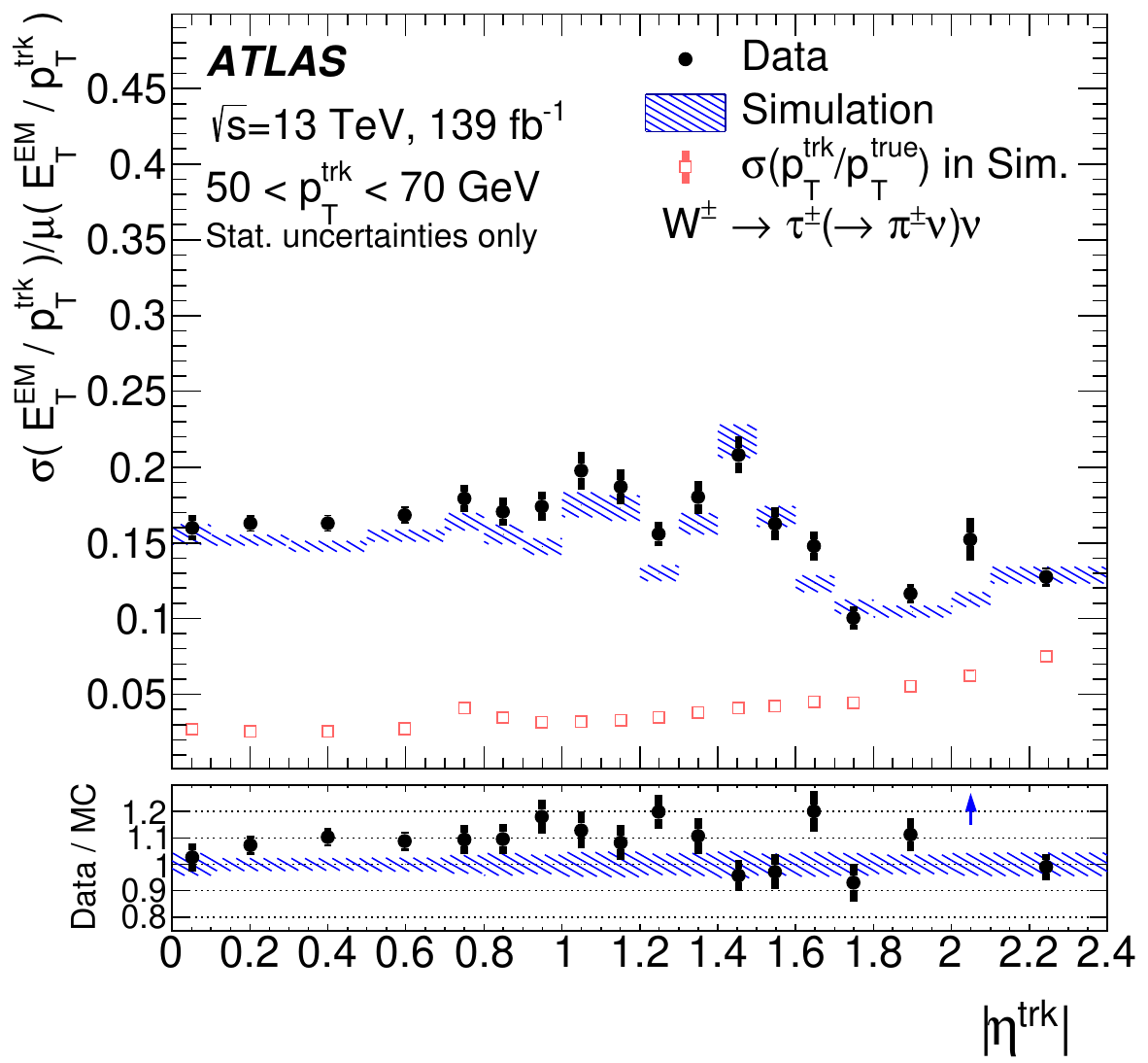}
\includegraphics[width=0.32\textwidth]{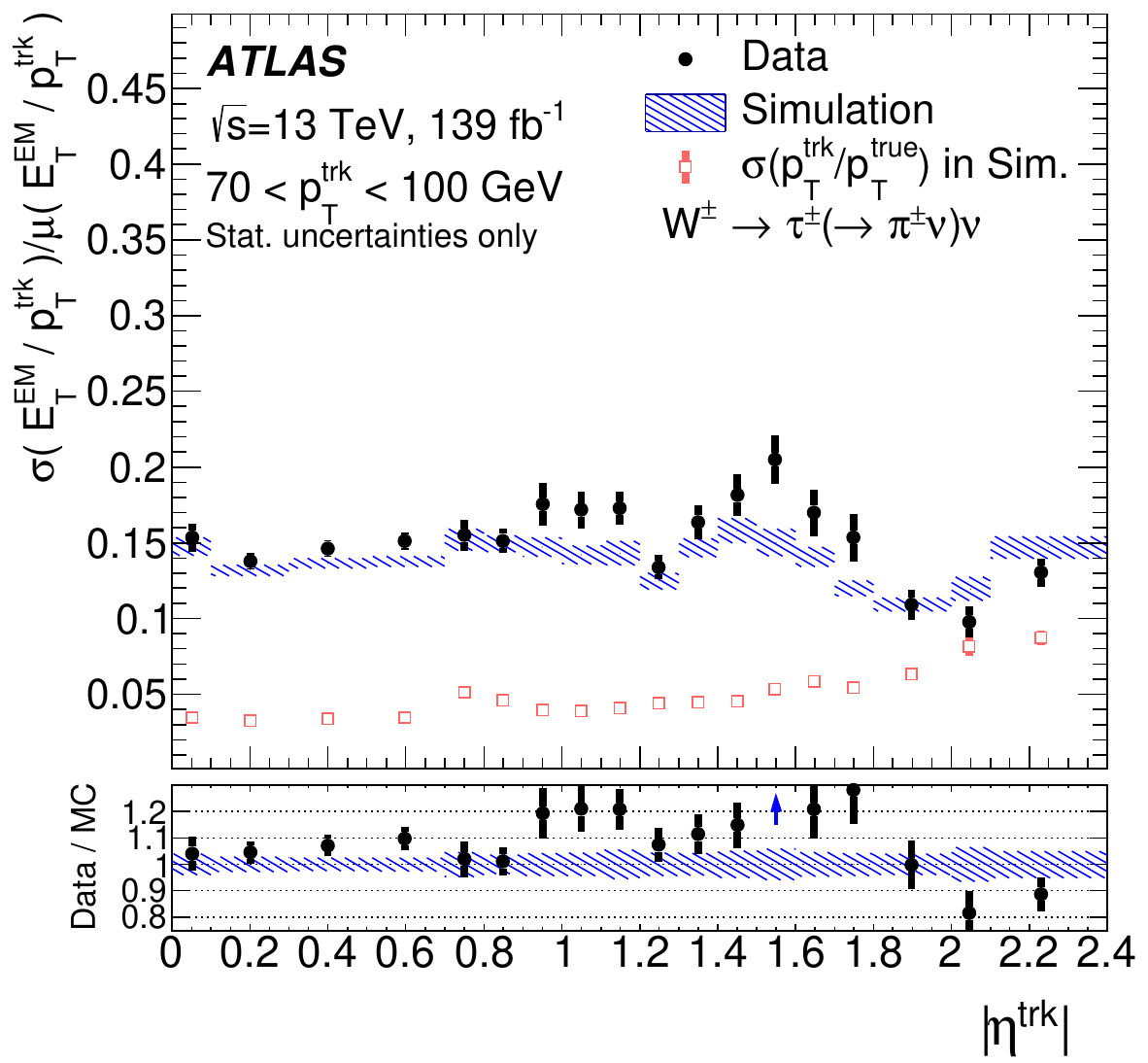}
\caption{The fitted width divided by the mean of the signal \EToverPT distribution as a function of the track $|\eta|$ in three \pTtrack ranges: $30<\pTtrack<50$~\GeV, $50<\pTtrack<70$~\GeV\ and $70<\pTtrack<100$~\GeV. Also shown is the fitted width of the track \pT divided by the generator-level \pT for true $\tau\rightarrow\pi^\pm\nu_\tau$ events in simulation to give an illustration of the contribution to the total width from the resolution of reconstructed tracks. The uncertainties shown are only those from the limited number of events in the dataset and simulated samples.}
\label{fig.resVsEta}
\end{figure}
 
% End of text imported from the .//sections/Width.tex input file
 
\section{Measurements of the longitudinal energy profile and late showering particles}
\label{sec:layers}
% The next lines are included from the .//sections/Layer.tex input file
The longitudinal segmentation of the calorimeter into layers can be used to gain insight into the cause of the discrepancies in the scale and resolution between data and simulation. Backgrounds can bias any measurements of average energies deposited in a single layer. In the electromagnetic calorimeter the main background, $\tau^\pm\rightarrow\pi^\pm+n\pi^0+\nu_\tau$, will result in significant biases as the $\pi^0\rightarrow\gamma\gamma$ showers will be contained within the electromagnetic calorimeter. Therefore, only the energy deposited in the hadronic calorimeter layers is considered, although these measurements are still slightly biased by this background through selection effects because the events are required to pass the hadronic energy fraction requirement. The hadronic calorimeter is less affected by \pu since low-momentum hadrons are often contained within the electromagnetic calorimeter in addition to the \pu particles that produce electromagnetic showers.
 
Figure~\ref{fig.layer} shows the average energy deposited in each layer of the hadronic calorimeter divided by the track momentum in the barrel and endcap regions as a function of the track \pT. Events are required to have $0.2<\EToverPT<1.1$ to select events across the full response and also minimise the contribution from backgrounds. An arithmetic mean of all the selected events is used because the distribution of the energy in an individual layer is not expected to have a Gaussian shape. It can be seen that in the barrel the simulation overestimates the relative energy in the hadronic calorimeter by 3-5\% across the \pT spectrum and this is the case in all layers, although most of the energy is contained within the first two layers. In the endcap region the amount of energy in the hadronic layers is slightly underestimated overall. This is driven by an underestimation of the energy in the first layer. Both of these features are in line with the results in Section~\ref{sec:results}; however, the magnitude of the difference in the energy deposited does not explain the size of the effect seen in the total response, indicating that there is mismodelling either in both the electromagnetic and hadronic calorimeters or in the length of the shower, or in a combination of these or other effects. The results are seen to be largely unaffected by \pu. These results can be used for the tuning of the detector simulation.
 
\begin{figure}
\centering
\includegraphics[width=0.45\textwidth]{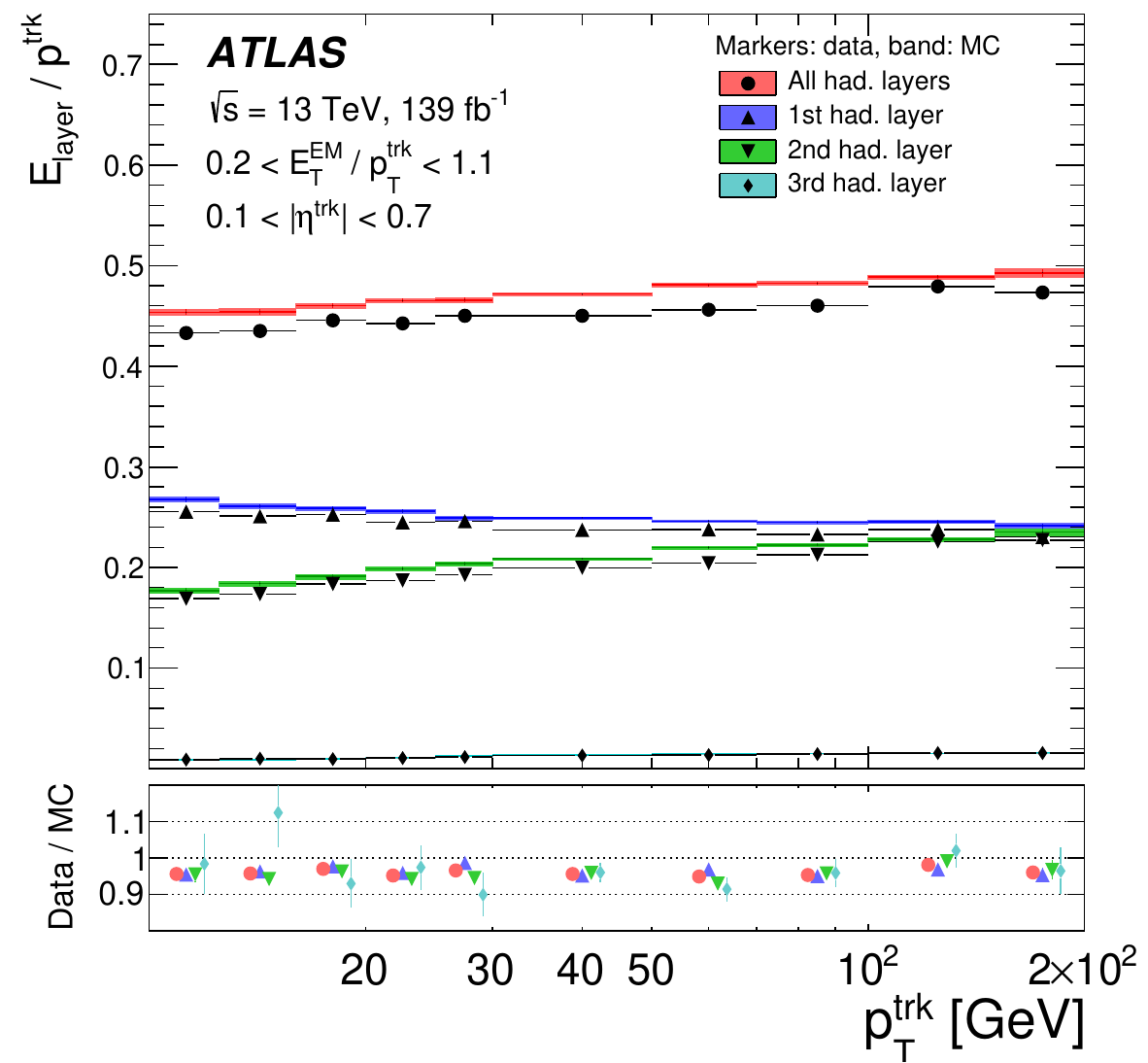}
\includegraphics[width=0.45\textwidth]{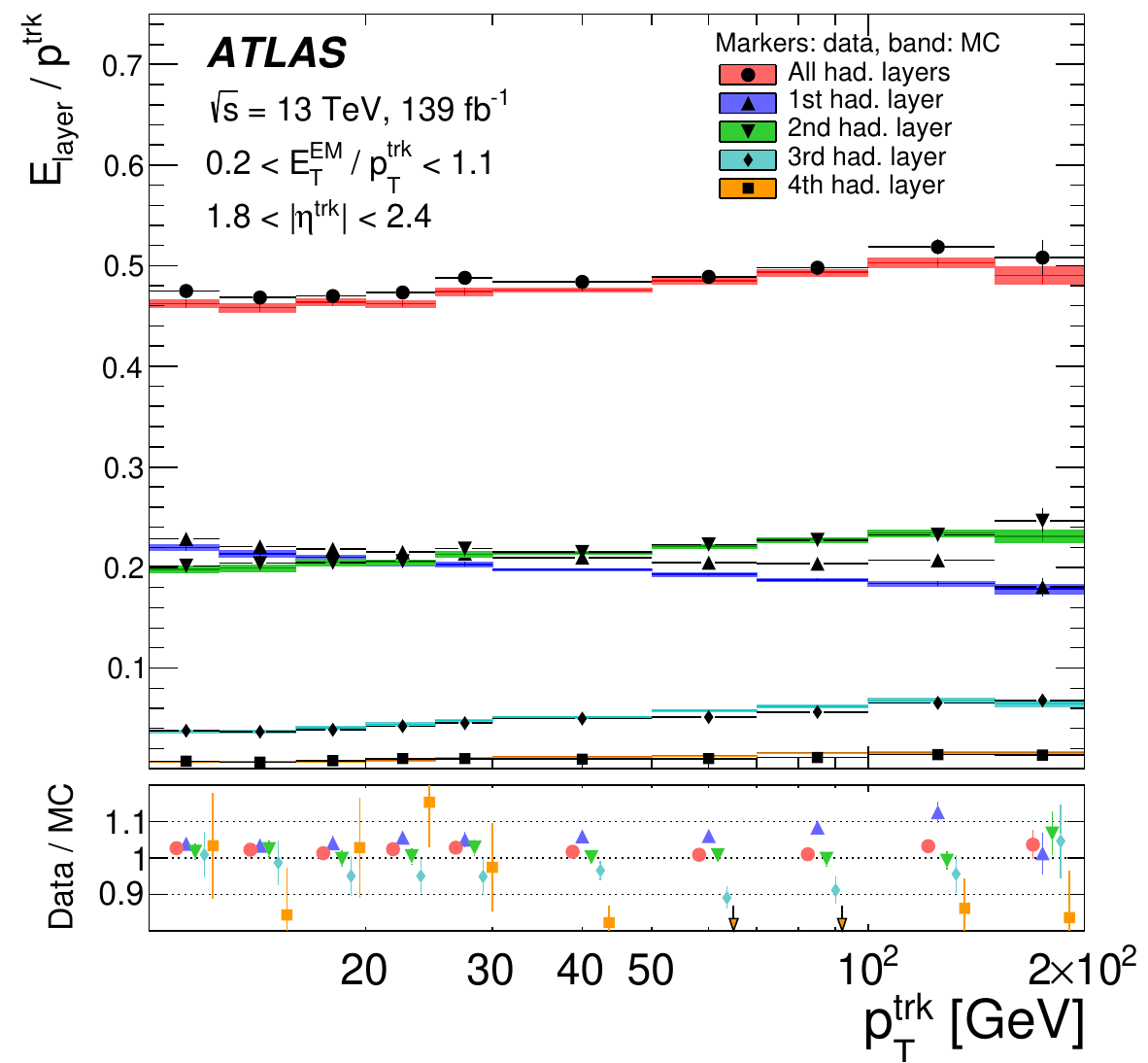}
\caption{The average energy within each hadronic layer normalised to the track momentum as a function of track \pT in the barrel (left) and endcaps (right). Also shown are the ratios of the sum of energies in all layers of the hadronic calorimeter to the track momentum. Data are required to have $0.2<\EToverPT<1.1$ to primarily select single-pion events. Shown in the lower panel is the ratio of data to simulation. The error bars and uncertainty band show the uncertainties from the limited number of events in the dataset and simulated samples.}
\label{fig.layer}
\end{figure}
 
A second method of using the longitudinal segmentation to probe the sources of the discrepancies is to select events where the pion traverses the electromagnetic calorimeter as a minimally ionising particle before showering in the hadronic calorimeter layers. Such events are characterised by having a very large fraction of their energy in the hadronic calorimeter. Selecting events in which more than 85\% of the summed cluster energy is in the hadronic calorimeters, $f^{\textrm{had}}>85\%$, allows the data-to-MC response ratio to be probed in the hadronic calorimeters. Figure~\ref{fig.MIP} shows the fitted response after applying this selection criterion in the barrel and endcap regions. In the barrel region the data-to-MC response ratio is seen to be similar to the inclusive response ratio; however, larger discrepancies are seen in the endcaps after this selection, with the difference between data and simulation increasing from ${\sim}4\%$ to ${\sim}8\%$. This suggests that the simulation models hadronic showers in the hadronic calorimeter less accurately in the endcaps than in the inclusive case.
 
\begin{figure}
\centering
\includegraphics[width=0.45\textwidth]{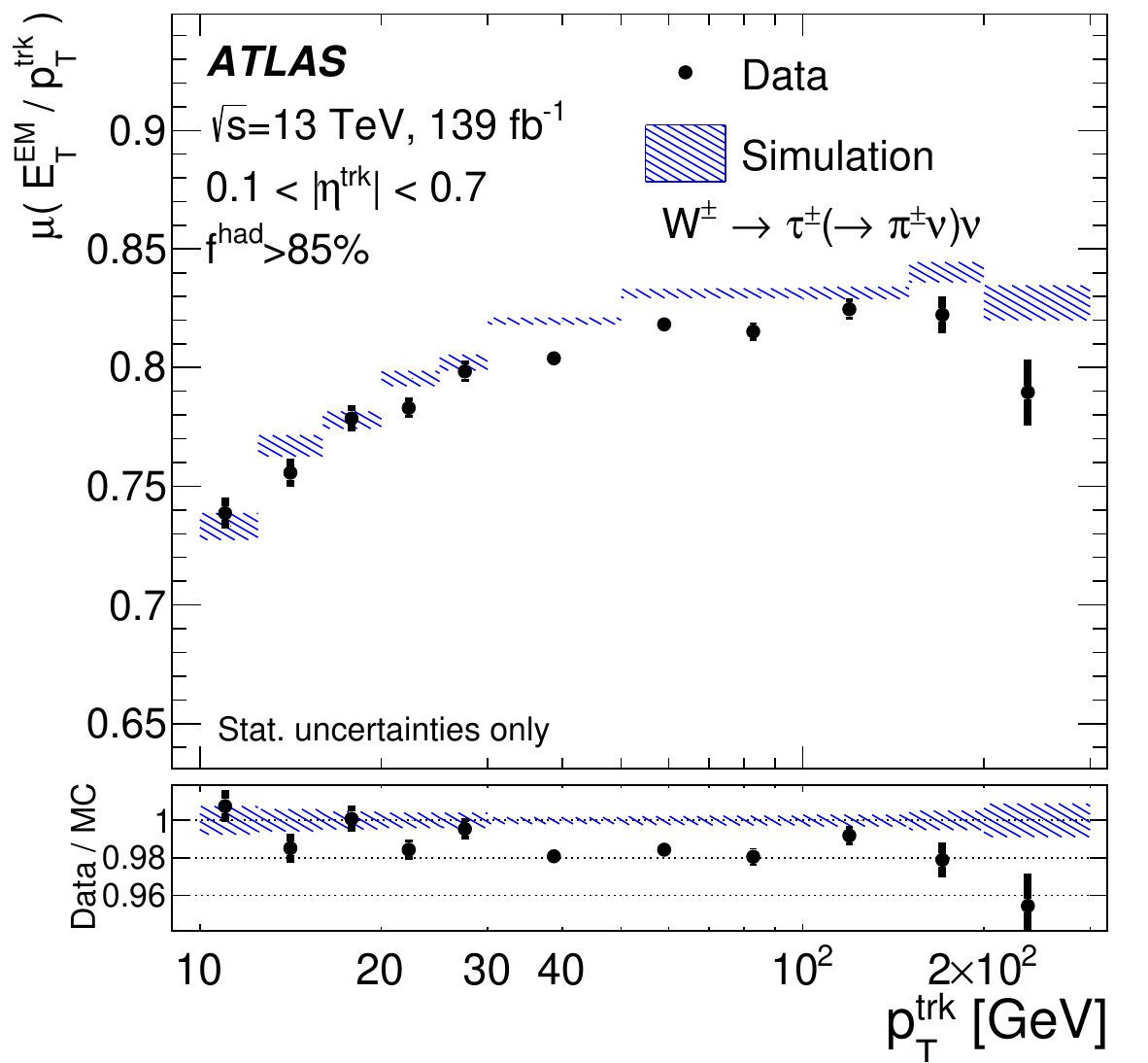}
\includegraphics[width=0.45\textwidth]{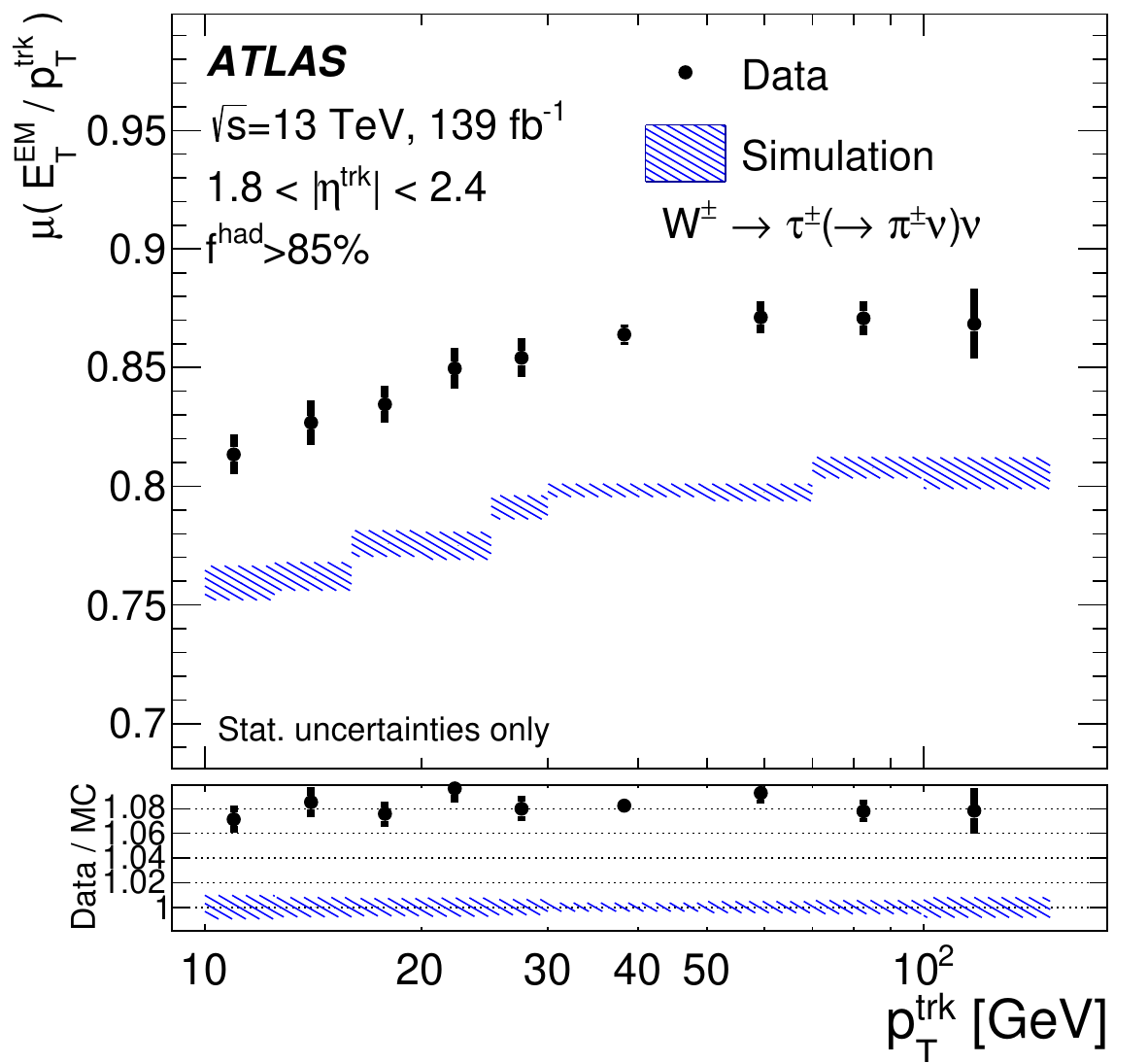}
\caption{The fitted response as a function of track \pT in the barrel (left) and endcaps (right) after requiring the fraction of the associated energy that is in the hadronic calorimeters to satisfy $f^{\textrm{had}}>85\%$. The error bars and uncertainty band show the uncertainties from the limited number of events in the dataset and simulated samples respectively.}
\label{fig.MIP}
\end{figure}
% End of text imported from the .//sections/Layer.tex input file
\newpage
\section{Conclusion}
\label{sec:conclusions}
% The next lines are included from the .//sections/Conclusion.tex input file
The energy response of the ATLAS calorimeter has been measured across a wide range of pion momenta with high precision using an innovative technique based on isolated charged pions from the process $W^\pm\rightarrow\tau^\pm(\rightarrow\pi^\pm+\nu_\tau)\nu_\tau$. The calorimeter response is observed to be overestimated by ${\sim}2\%$ in the central region and underestimated by ${\sim}4\%$ in the endcaps in the ATLAS simulation. This supports the observations in the measurement of the jet energy scale, which show a similar structure across the calorimeter \cite{JETM-2018-05}. The precision of the measurement of the response is ${\lesssim}1\%$ for $15<\pT<185$~\GeV\ ($20<\pT<100$~\GeV) in the central (endcap) regions, and is ${<}0.6\%$ in the most precise region, $20<\pT<120$~\GeV\ in the barrel.
 
The width of the \EToverPT distribution has also been investigated. This measurement convolves the tracking and calorimeter resolutions. For this quantity the simulation is found to agree with the data to within 15\%; however, the data are generally observed to have a wider distribution. Finally, the average energy in the layers of the hadronic calorimeter was probed as a method of investigating the cause of the discrepancies between the simulation and the data. These measurements will be used in the tuning of the ATLAS simulation to further improve the hadron shower modelling and detector geometry.

This powerful new method of measuring the hadronic energy response of the ATLAS calorimeter can achieve a precision of ${\lesssim}1\%$ for the response of the calorimeter to charged pions. This detailed understanding of the response of the calorimeter can be used in gaining better understanding of the jet energy scale, its uncertainty for the highest-\pT jets, the energy scale of hadronically decaying $\tau$-leptons, and in measuring the properties of jets.

% End of text imported from the .//sections/Conclusion.tex input file
 
\section*{Acknowledgements}
% The next lines are included from the .//acknowledgements/Acknowledgements.tex input file

We thank CERN for the very successful operation of the LHC, as well as the
support staff from our institutions without whom ATLAS could not be
operated efficiently.
 
We acknowledge the support of
ANPCyT, Argentina;
YerPhI, Armenia;
ARC, Australia;
BMWFW and FWF, Austria;
ANAS, Azerbaijan;
SSTC, Belarus;
CNPq and FAPESP, Brazil;
NSERC, NRC and CFI, Canada;
CERN;
ANID, Chile;
CAS, MOST and NSFC, China;
Minciencias, Colombia;
MEYS CR, Czech Republic;
DNRF and DNSRC, Denmark;
IN2P3-CNRS and CEA-DRF/IRFU, France;
SRNSFG, Georgia;
BMBF, HGF and MPG, Germany;
GSRI, Greece;
RGC and Hong Kong SAR, China;
ISF and Benoziyo Center, Israel;
INFN, Italy;
MEXT and JSPS, Japan;
CNRST, Morocco;
NWO, Netherlands;
RCN, Norway;
MEiN, Poland;
FCT, Portugal;
MNE/IFA, Romania;
JINR;
MES of Russia and NRC KI, Russian Federation;
MESTD, Serbia;
MSSR, Slovakia;
ARRS and MIZ\v{S}, Slovenia;
DSI/NRF, South Africa;
MICINN, Spain;
SRC and Wallenberg Foundation, Sweden;
SERI, SNSF and Cantons of Bern and Geneva, Switzerland;
MOST, Taiwan;
TAEK, Turkey;
STFC, United Kingdom;
DOE and NSF, United States of America.
In addition, individual groups and members have received support from
BCKDF, CANARIE, Compute Canada and CRC, Canada;
COST, ERC, ERDF, Horizon 2020 and Marie Sk{\l}odowska-Curie Actions, European Union;
Investissements d'Avenir Labex, Investissements d'Avenir Idex and ANR, France;
DFG and AvH Foundation, Germany;
Herakleitos, Thales and Aristeia programmes co-financed by EU-ESF and the Greek NSRF, Greece;
BSF-NSF and GIF, Israel;
Norwegian Financial Mechanism 2014-2021, Norway;
NCN and NAWA, Poland;
La Caixa Banking Foundation, CERCA Programme Generalitat de Catalunya and PROMETEO and GenT Programmes Generalitat Valenciana, Spain;
G\"{o}ran Gustafssons Stiftelse, Sweden;
The Royal Society and Leverhulme Trust, United Kingdom.
 
The crucial computing support from all WLCG partners is acknowledged gratefully, in particular from CERN, the ATLAS Tier-1 facilities at TRIUMF (Canada), NDGF (Denmark, Norway, Sweden), CC-IN2P3 (France), KIT/GridKA (Germany), INFN-CNAF (Italy), NL-T1 (Netherlands), PIC (Spain), ASGC (Taiwan), RAL (UK) and BNL (USA), the Tier-2 facilities worldwide and large non-WLCG resource providers. Major contributors of computing resources are listed in Ref.~\cite{ATL-SOFT-PUB-2021-003}.
 
% End of text imported from the .//acknowledgements/Acknowledgements.tex input file
 
\clearpage
 
\printbibliography

\clearpage
% ATLAS Collaboration author list
% Reference date of JETM-2018-08 is 2021-06-18
% Author list last updated on date 02-SEP-22
% Data extracted on 02-Sep-2022 for paper reference JETM-2018-08
% at 11:55am
 
\begin{flushleft}
\hypersetup{urlcolor=black}
{\Large The ATLAS Collaboration}

\bigskip

\AtlasOrcid[0000-0002-6665-4934]{G.~Aad}$^\textrm{\scriptsize 98}$,    
\AtlasOrcid[0000-0002-5888-2734]{B.~Abbott}$^\textrm{\scriptsize 123}$,    
\AtlasOrcid[0000-0002-7248-3203]{D.C.~Abbott}$^\textrm{\scriptsize 99}$,    
\AtlasOrcid[0000-0002-2788-3822]{A.~Abed~Abud}$^\textrm{\scriptsize 34}$,    
\AtlasOrcid[0000-0002-1002-1652]{K.~Abeling}$^\textrm{\scriptsize 51}$,    
\AtlasOrcid[0000-0002-2987-4006]{D.K.~Abhayasinghe}$^\textrm{\scriptsize 91}$,    
\AtlasOrcid[0000-0002-8496-9294]{S.H.~Abidi}$^\textrm{\scriptsize 27}$,    
\AtlasOrcid[0000-0002-9987-2292]{A.~Aboulhorma}$^\textrm{\scriptsize 33e}$,    
\AtlasOrcid[0000-0001-5329-6640]{H.~Abramowicz}$^\textrm{\scriptsize 156}$,    
\AtlasOrcid[0000-0002-1599-2896]{H.~Abreu}$^\textrm{\scriptsize 155}$
% ...
\bigskip
\\

% ...
$^{27}$Physics Department, Brookhaven National Laboratory, Upton NY; United States of America.\\
% ...
$^{33}$$^{(a)}$Facult\'e des Sciences Ain Chock, R\'eseau Universitaire de Physique des Hautes Energies - Universit\'e Hassan II, Casablanca;$^{(b)}$Facult\'{e} des Sciences, Universit\'{e} Ibn-Tofail, K\'{e}nitra;$^{(c)}$Facult\'e des Sciences Semlalia, Universit\'e Cadi Ayyad, LPHEA-Marrakech;$^{(d)}$LPMR, Facult\'e des Sciences, Universit\'e Mohamed Premier, Oujda;$^{(e)}$Facult\'e des sciences, Universit\'e Mohammed V, Rabat; Morocco.\\
$^{34}$CERN, Geneva; Switzerland.\\
% ...
$^{51}$II. Physikalisches Institut, Georg-August-Universit\"{a}t G\"ottingen, G\"ottingen; Germany.\\
% ...
$^{91}$Department of Physics, Royal Holloway University of London, Egham; United Kingdom.\\
% ...
$^{98}$CPPM, Aix-Marseille Universit\'e, CNRS/IN2P3, Marseille; France.\\
$^{99}$Department of Physics, University of Massachusetts, Amherst MA; United States of America.\\
% ...
$^{123}$Homer L. Dodge Department of Physics and Astronomy, University of Oklahoma, Norman OK; United States of America.\\
% ...
$^{155}$Department of Physics, Technion, Israel Institute of Technology, Haifa; Israel.\\
$^{156}$Raymond and Beverly Sackler School of Physics and Astronomy, Tel Aviv University, Tel Aviv; Israel.\\
% ...
$^{a}$ Also at Borough of Manhattan Community College, City University of New York, New York NY; United States of America.\\
$^{b}$ Also at Bruno Kessler Foundation, Trento; Italy.\\
$^{c}$ Also at Center for High Energy Physics, Peking University; China.\\
$^{d}$ Also at Centro Studi e Ricerche Enrico Fermi; Italy.\\
$^{e}$ Also at CERN, Geneva; Switzerland.\\
% ...

\end{flushleft}

% Created with Glance <Atlas.Glance@cern.ch>

\end{document}